\documentclass{aa}  

\usepackage{graphicx}
\usepackage{comment}
\usepackage{txfonts}
\usepackage{natbib}
\bibpunct{(}{)}{;}{a}{}{,} 

\usepackage[graphicx]{realboxes}

\usepackage[colorlinks=true,linkcolor=blue,citecolor=blue,urlcolor=blue]{hyperref}
\newcommand{\ie}{i.e.}
\newcommand{\eg}{e.g.}
\newcommand{\JKTEBOP}{\textsc{jktebop}}

\newcommand{\Kepler}{\textit{Kepler}}

\newcommand{\Gaia}{\textit{Gaia}}
\newcommand{\spd}{\textsc{spd}}

\newcommand{\Pin}{$P_\mathrm{A}$}
\newcommand{\Pout}{$P_\mathrm{AB}$}
\newcommand{\Toin}{$T^\mathrm{A}_\mathrm{0}$}

\newcommand{\eout}{$e_\mathrm{AB}$}
\newcommand{\ein}{$e_\mathrm{A}$}
\newcommand{\incout}{$i_\mathrm{AB}$}
\newcommand{\incin}{$i_\mathrm{A}$}
\newcommand{\incmut}{$i_\mathrm{mut}$}

\newcommand{\smain}{$a_\mathrm{A}$}

\newcommand{\logg}[1]{$logg_\mathrm{#1}$ }

\newcommand{\vsin}[1]{$v_{#1}\sin{i}$}

\newcommand{\rebound}{\textsc{rebound}}

\begin{document}

   \title{Spectroscopy of eclipsing compact hierarchical triples}

   \subtitle{I. Low-mass double-lined and triple-lined systems}

   \author{Ayush Moharana
          \inst{1}
          \and
          K. G. He{\l}miniak\inst{1}
          \and
          F. Marcadon\inst{2}
          \and
          T. Pawar\inst{1}
          \and
          G. Pawar\inst{1}
          \and
          M. Konacki\inst{1,3}
          \and
          A. Jord\'{a}n\inst{4,5,6} 
          \and
          R.~Brahm\inst{4,5,6} 
          \and
          N. Espinoza\inst{7} 
          }

   \institute{Nicolaus Copernicus Astronomical Center, Polish Academy of Sciences, ul.                   Rabia\'{n}ska 8, 87-100 Toru\'{n}, Poland 
   \and
   Villanova University, Dept.\ of Astrophysics and Planetary Sciences, 800 East Lancaster Avenue, Villanova, PA 19085, USA 
    \and
     Nicolaus Copernicus Astronomical Center, Polish Academy of Sciences, ul. Bartycka 18, 00-716 Warszawa, Poland
    \and
    Facultad de Ingenier\'ia y Ciencias, Universidad Adolfo Ib\'a\~nez, Av.\ Diagonal las Torres 2640, Pe\~nalol\'en, Santiago, Chile
    \and
    Millennium Institute for Astrophysics, Monse\~nor Nuncio S\'otero Sanz 100, Of. 104, Providencia, Santiago, Chile
    \and
    Data Observatory Foundation, Eliodoro Y\'a\~nez 2990, Providencia, Santiago, Chile
    \and
    Space Telescope Science Institute, 3700 San Martin Dr., Baltimore 21218, MD, USA
             }

   \date{Received September 15, 1996; accepted March 16, 1997}

  \abstract{Eclipsing compact hierarchical triples (CHTs) are systems in which a tertiary star orbits an eclipsing binary (EB) in an orbit of fewer than 1000 days. In a CHT, all three stars exist in a space that is less than 5 AU in radius. A low-mass CHT is an interesting case through which we can understand the formation of multiple stars and planets at such small scales.}
   {In this study, we combine spectroscopy and photometry to estimate the orbital, stellar, and atmospheric parameters of stars in a sample of CHTs. Using the complete set of parameters, we aim to constrain the metallicity and age of the systems.}
   {We used time-series spectroscopy to obtain radial velocities (RVs) and disentangled spectra. Using RV modelling, EB light curve modelling, and spectral analysis, we estimated the metallicities and temperatures. Using isochrone fitting, we constrained the ages of the system.  We then combined observations of masses, outer eccentricities ($e_2$), orbital periods, and age estimates of the systems from the literature. We compared the distributions of $e_2$, and the tertiary mass ratio, $q_3 = M_3/(M_1+M_2)$, for three different metallicity ranges and two age ranges.}
   {We have estimated the masses, radii, temperatures, metallicities, and ages of 12 stars in four CHTs. The CHT CD-32 6459 shows signs of von Zeipel-Lidov-Kozai oscillations, while CD-62 1257 can evolve to form a triple common envelope. The rest of the CHTs are old and have an M-dwarf tertiary. We find that the $q_3$ distribution for CHTs with sub-solar metallicity has a uniform distribution but the systems with solar and above-solar metallicity peak between 0.5 and 1. When dividing them according to their ages, we find the $q_3$ of old systems to be around 0.5. The eccentricity, $e_2$, favours a value of around 0.3 irrespective of metallicity or age. The distributions of $q_3$ and $e_2$ resemble the distributions of the mass ratio and eccentricity of close field binaries.}
   {}

   \keywords{binaries: eclipsing --  binaries: spectroscopic -- stars: fundamental parameters -- stars: evolution --  stars: formation -- stars: individual: CD-32\, 6459;  CD-62\, 1257; CD-58\, 963; BD+11\, 359   }

   \maketitle
%
\section{Introduction}
Stellar multiplicity is a well-established phenomenon \citep{multiplicity}. While thousands of binaries and multiples were observed in the 19th century \citep{Herschel187410Kstars}, \cite{Harrington1972CeMec} was one of the first to use the observations to understand multiple stars as a separate population. Later, with better samples and improved theories, we started having a better understanding of triple (and multiple) stars around the new millennium \citep{Eggleton1995,mardlingaasreth}. 

The previous and present decades have seen a revolution in observing and understanding multiple stars. While it was radial velocity (RV) surveys \citep{tokostats2004} that helped us to identify different hierarchies of multiple-star systems, most of the new detections have been from photometric surveys like the Optical Gravitational Lensing Experiment (OGLE; \citealt{ogle1}) and the All Sky Automated Survey (ASAS; \citealt{asas}). Recently, space-based surveys like \Kepler{} \citep{kepler} and the Transiting Exoplanet Survey Satellite (TESS; \citealt{TESSricker}) have been revolutionary in detecting and characterising binaries, triples, and multiple stars. 

Triples, especially, have been the subject of renewed interest recently. They are emerging as possible explanations for several problems in stellar astrophysics. This includes triple dynamics as an explanation for asymmetry of planetary nebula \citep{pnebfromtrip} formation of Thorne-Zytków objects \citep{eisnercht}, blue stragglers \citep{peretsfab}, recurrent novae \citep{novaetriple}, and Type Ia supernovae \citep{naozobliqtrip}. 

A special class of triple stars, compact hierarchical triples (CHTs), have seen increased incidence rates, which is surprising as they were considered rare before \citep{tokostats2004}. Compact hierarchical triples are hierarchical systems in which the tertiary orbits an inner binary with an orbital period of less than 1000 days \citep{borkoCHTreview}. This, in principle, can cause dynamic changes in these systems that can be characterised by a few years of observation. If we have an eclipsing binary (EB) as the inner binary, we can extract the parameters of each component in the system. This has led to eclipse timing programs helping us find hundreds of new CHTs from different EB catalogues \citep{borkovitskeptrip,ogletrip,TESStriplescvz}.

Estimating the precise stellar and orbital parameters of CHTs opens up an avenue through which to study stellar evolution coupled with dynamical evolution. Further, using the distribution of orbital parameters and masses, we can understand star formation at scales at which planet formation usually occurs ($\leq5$ AU).  While triply eclipsing systems (E3CHTs) can provide ultra-precise mass and radius measurements \citep{Borko2019}, using time-series spectroscopy of doubly eclipsing systems (E2CHTs) can also help us fill the same parameter space \citep{moharanacht}.

In this paper, we present the total parameters of four E2CHTs using high-resolution spectroscopy coupled with TESS{} photometry. Two of these systems, CD-58 963 (hereby CD58) and  BD+11 359 (hereby BD11), are spectroscopic double-lined (ST2) systems, while CD-62 1257 (hereafter CD62) and CD-32 6459 (hereby CD32) are triple-lined (ST3) systems. Three of these systems (CD32, CD62, and BD11) are newly identified CHTs. 

CD32 was first classified as an eccentric EB by \citep{shivvers2014} using observations from the All Sky Automated Survey (ASAS; \citealt{asas0-6southern}). \cite{Kim2018ETV} was the first to observe variation in eclipse times in the system, but they did not find any tertiary star signatures. While these works provide accurate estimates of the period and eccentricity, we provide the first measurements of stellar parameters of the EB stars and the tertiary companion. Though CD32 is not a CHT by the strictest of definitions (outer period of $\sim$1300d), we try to see if they are any different. 

CD62 was first flagged as an EB in the first TESS{} EB catalogue \citep{TESSEBPrsa2022}. The first light curve (LC) analysis was done by \cite{cd62firstpar}, but with the assumption that the system is a binary. This affected the estimated parameters as the tertiary in the system contributes a significant amount of third light ($\sim$50\%), which was assumed to be zero in their analysis. 

CD58 was first identified as a multiple system with EB by \cite{2020borko2tt}. It was one of the first CHTs discovered with TESS{}, as it is in the continuous viewing zone (CVZ) and has been observed since the first year of TESS{}. With eclipse timing from Wide-Angle Search for Planets \citep[WASP;][]{pollaco_wasp}, \cite{2020borko2tt} showed that CD58 is a hierarchical quadruple on a wide 2661 d orbit around the CHT. We do not find any signs of the quadruple but the point to note is that we do not have good coverage over the quadruple period. CD58 is the tightest CHT in this sample, with an outer-orbit period of 76.32 d. Such systems are considered very rare, as it is difficult to survive the migration evolution from early formation \citep{tokomoe2020}. 

BD11 was identified as an EB in the ASAS survey and its first spectroscopic and LC solution was given by \cite{krisasas1}. Later, \citet{bachestan} used it as a test object for their BACHES spectrograph and also presented an initial orbital solution. None of those works describes BD11 as a triple.

\section{Observations}
\label{sec:Observations}
\subsection{Photometry}
All of the targets were observed by TESS\footnote{Through Guest Investigator (GI) programs G011083, G04047, G05078 (PI: He{\l}miniak), and G05003 (PI: Pr\v{s}a). CD58 was also included in the TESS{} Core Target Sample (CTL) during Cycle 1.} for at least two sectors. For our work, we chose the best sectors considering (i) minimal cadence, (ii) low stellar activity or out-of-eclipse variations, and (iii) long-term coverage to detect eclipse depth variations (EDVs), if any. All the chosen sectors had a minimal cadence of two minutes. CD58 has been studied before, using sectors 1 to 12. We analysed the system with new observations from sectors 62 to 69.
We extracted LCs using the \textsc{lightkurve}\footnote{\url{https://docs.lightkurve.org/}} package. We extracted the photometry using the standard pipeline aperture.  

While CD32 (TIC 24972851) and CD58 (TIC 220397947) are well isolated in the TESS{} frames, CD62 (TIC 387107961) and BD11 (TIC 408834852) have close-by stars. We checked for any contaminant signature (\eg pulsations, eclipses, or transits) but did not find any, and therefore used the pipeline photometry.

The detrending was done with \textsc{wotan}\footnote{\url{https://github.com/hippke/wotan}} \citep{wotan}. We used the \texttt{bi-weight} de-trending method in a window 0.5-3 times the orbital period (depending on the trends). The normalised output from \textsc{wotan} was then converted to the magnitude scale by using zero-points that adjusted the out-of-eclipse magnitude to the TESS{} magnitude registered in the TESS{} catalogue.  

\subsection{Spectroscopy} 
The spectroscopy was obtained from a set of high-resolution spectrographs that includes a Fibre-fed Extended Range Optical Spectrograph (FEROS; R$\sim$48,000) at the MPG/ESO 2.2m telescope in La Silla \citep{feros}, CHIRON (R$\sim$28,000 in the fiber mode) at the CTIO 1.5m telescope in Cerro Tololo \citep{Tokovinin2013}, CORALIE (R$\sim$70,000) at the 1.2m Euler telescope in La Silla \citep{coralie}, the High-Resolution Spectrograph (HRS; R$\sim$67,000) at the 9.2m SALT in Sutherland \citep{hrs3}, and the High Accuracy Radial velocity Planet Searcher (HARPS; R$\sim$115,000) at the 3.6m ESO telescope in La Silla \citep{harps}. 
Additionally, for BD11 we also used RV measurements from \citet{krisasas1} that were based on data obtained with the University College London Echelle Spectrograph (UCLES) at the 3.9m AAT in Siding Spring Observatory. We did not use BACHES data from \citet{bachestan}, as they are of significantly lower quality.

The CORALIE and FEROS spectrographs both work in a simultaneous object-calibration manner. Spectra were reduced with the dedicated python-based pipeline \citep{2014AJ....148...29J,brahm_ceres}, optimised to derive high-precision RVs, which also performs barycentric corrections. The pipeline reduces CORALIE spectra to 70 rows spanning from 3840 to 6900~\AA, of which we used only 45 rows (4400--6500~\AA), to avoid the broad H$_\alpha$ line and the blue part with a very low signal. For FEROS, the output was reduced to 21 rows covering 4115-6519~\AA, of which we used 20 (4135--6500~\AA).

The CHIRON spectra were reduced with the pipeline developed at Yale University \citep{Tokovinin2013}. Wavelength calibration was based on ThAr lamp exposures taken just before the science observation. Barycentric corrections are not applied by the pipeline; thus, we calculated them ourselves under IRAF\footnote{IRAF is distributed by the National Optical Astronomy Observatory (NOAO), which is operated by the Association of Universities for Research in Astronomy (AURA) under a cooperative agreement with the National Science Foundation. \url{http://iraf.noao.edu/}} with the {\it bcvcor} task.

The HRS spectra were obtained by the long-term programme 2021-2-MLT-006 (PI: Moharana), which focused on the spectroscopic monitoring of CHTs. They were made available after reduction with the MIDAS HRS pipeline \citep{saltmidas1,saltmidas2}. While the products included spectra in the blue arm (370-550 nm) and the red arm (550-890 nm), we used the blue arm to avoid contamination by the static telluric lines. The barycentric correction was also done with {\it bcvcor}.

The HARPS data were reduced on site, including wavelength calibration and barycentric correction, with ESO's data reduction system (DRS). They are available through the ESO Archive.

\begin{table}
    \centering
    \caption{All the spectroscopic observations used for RV extraction.}
    \label{tab:spec_obs}    \begin{tabular}{ccccc}
    \hline
         Targets& CD32 & CD62 & CD58 & BD11 \\
         \hline
         \hline
         FEROS  & 4 & 8 & - & 11 \\           
         CHIRON &15 &12 & - & -\\    
         CORALIE& 1 & 6 & - & 5\\            
         HRS    & 4 & 8 &13 & -\\            
        HARPS   & - &-  &-  & 2\\
        UCLES   & - &-  &-  & 5$^a$\\          
        \hline                              
          Total &24 &34 &13 & 23\\
         \hline
    \end{tabular}
\\$^a$ From \citet{krisasas1}
\end{table}

\section{Analysis}

\subsection{Radial velocity extraction and fitting}
\label{sec:rvextract}

The RVs were calculated with a TODCOR method \citep{zuckermazeh}, with synthetic spectra computed with ATLAS9 code as templates. Measurement errors were calculated with a bootstrap approach, and used for weighting the measurements during the orbital fit, as they are sensitive to the signal-to-noise ratio (S/N) of the spectra and rotational broadening of the lines. Though this code is optimised for double-lined spectroscopic binaries (SB2) and provides velocities for two stars ($u_1, u_2$), it can still be used for ST3 as well. In an ST3, the tertiary's velocities were found from a local maximum, where $u_1$ was set for the tertiary, and $u_2$ for the brighter component of the eclipsing pair. 

The orbital solutions were found using our procedure called {\sc v2fit} \citep{v2fit}. It applies a Levenberg-Marquardt minimisation scheme to find orbital parameters of a double-Keplerian orbit, which can optionally be perturbed by several effects, like a circumbinary body. We fitted for the binary period ($P_\mathrm{A}$), time of periastron passage of the inner orbit ($T_{pA}$), inner eccentricity (\ein), argument of periastron ($\omega$), semi-amplitudes of the EB ($K_{Aa}$,$K_{Ab}$), projection of the semi-major axis ($a_{A} \sin{i_1}$), and systemic velocity ($ \gamma_{A} $).
The fitting follows the procedure defined in  \cite{krishides2017}, \cite{moharanacht}, and references therein.

In addition, for each observation for which three sets of lines were sufficiently separated, we also calculated the systemic velocities, $\gamma(t_i)$, of the inner pair using the formula
\begin{equation}
\gamma(t_i) = \frac{v_1(t_i)+ q v_2(t_i)}{1+q},
\label{eq:q_gam}
\end{equation}
where $v_{1,2}(t_i)$ are the measured RVs of the inner binary, and $q$ is its mass ratio, found from the RV fit with a circumbinary perturbation. With these values as the centre-of-mass (COM) RVs of the binary, and RVs of the tertiary component, we can treat the long-period outer orbit as a spectroscopic double-lined system, and independently look for its parameters. The final values of \Pout, $K_\mathrm{out}$, \eout{}, and so on, come from such fits. For the triple spectroscopic double-lined systems (ST2), this is the only way of estimating the orbital parameters of the third body and its projected mass ($M_\mathrm{B}\sin{i_\mathrm{AB}}$).

\begin{figure*}
    \centering
    \includegraphics[width=0.45\textwidth]{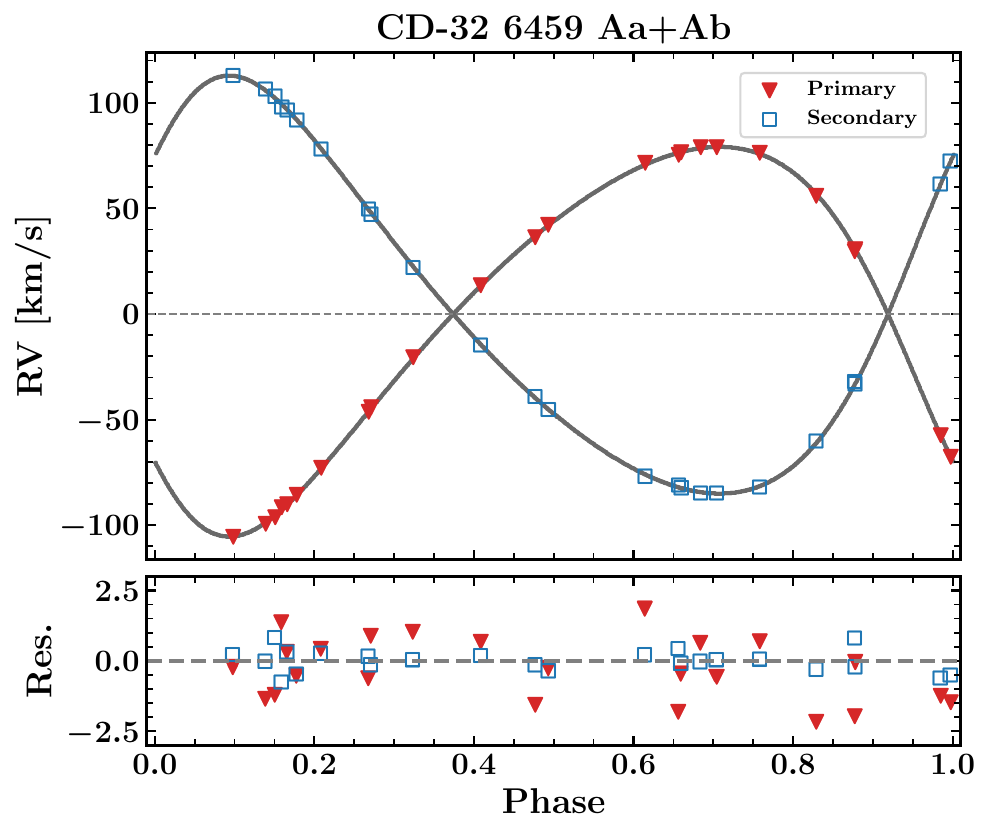}
    \includegraphics[width=0.45\textwidth]{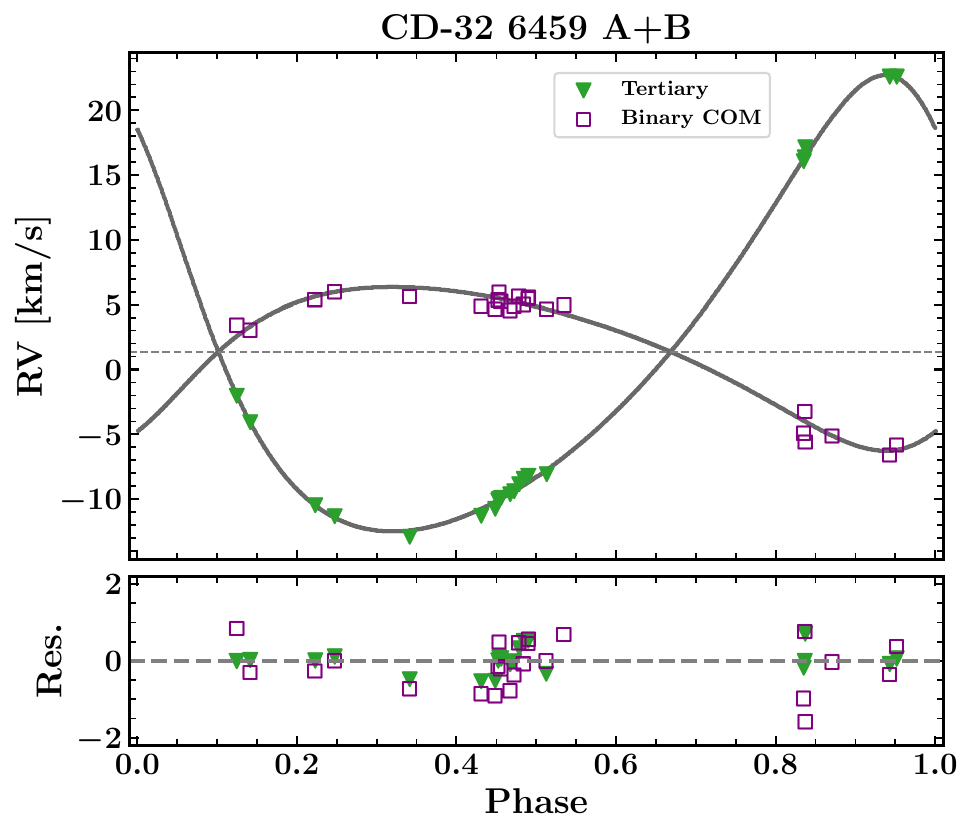}
    \includegraphics[width=0.45\textwidth]{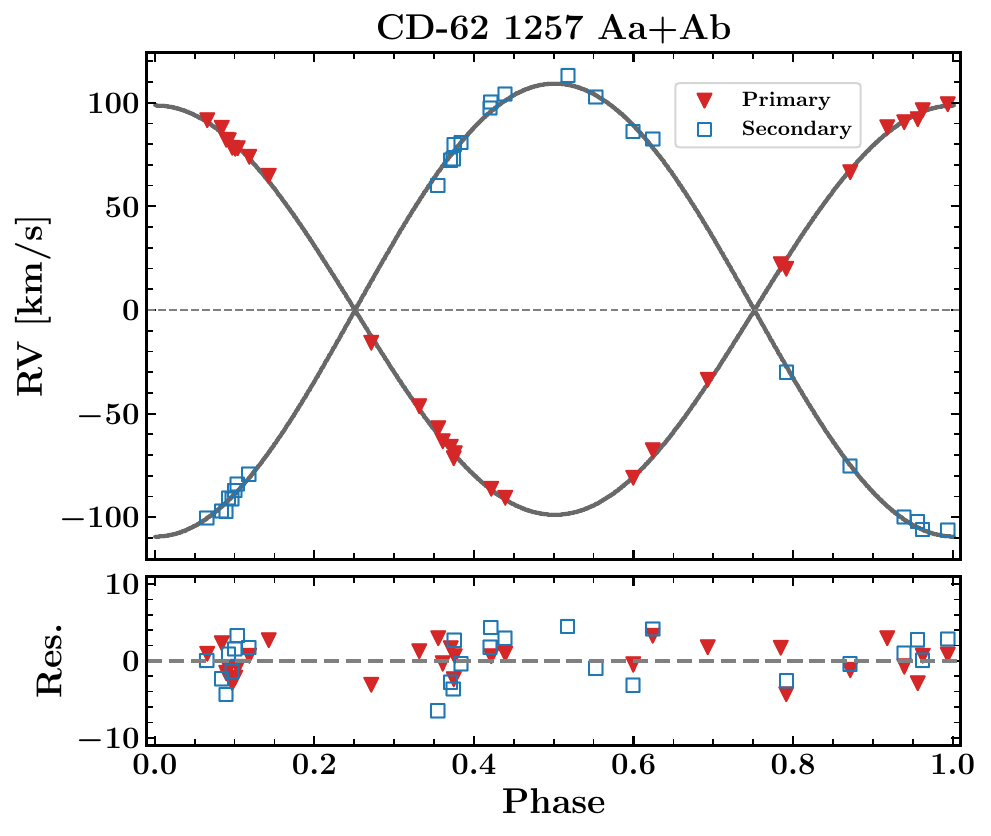}
    \includegraphics[width=0.45\textwidth]{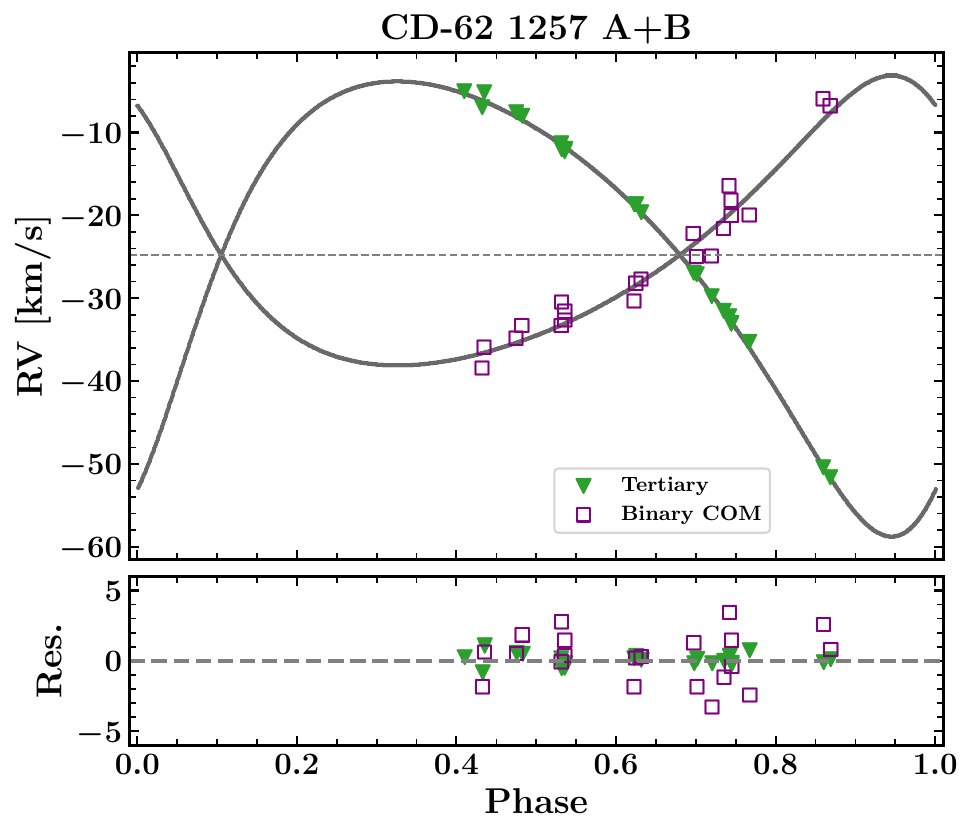}
   
\caption{RVs and orbital solutions of the ST3 systems CD32 (top) and CD62 (bottom). The left panels present the inner binary Aa+Ab corrected for its COM. Primary components are marked with red triangles and secondary ones with blue squares. The right panels present the outer orbit, with the inner binary's COM (calculated with \autoref{eq:q_gam}) represented by purple squares and the tertiary's RVs by green triangles. The dashed line marks the $\gamma$ velocity of the whole system.}
 \label{fig:st3rvfit}
\end{figure*}

\begin{figure*}
    \centering
    \includegraphics[width=0.45\textwidth]{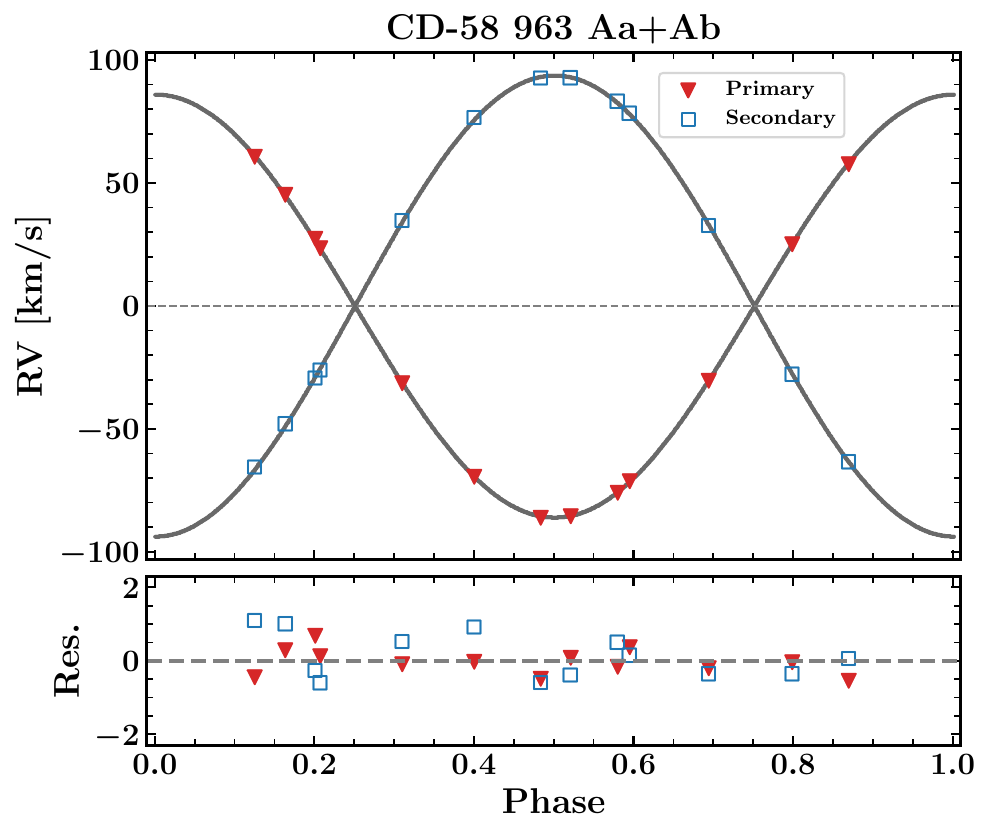}
    \includegraphics[width=0.45\textwidth]{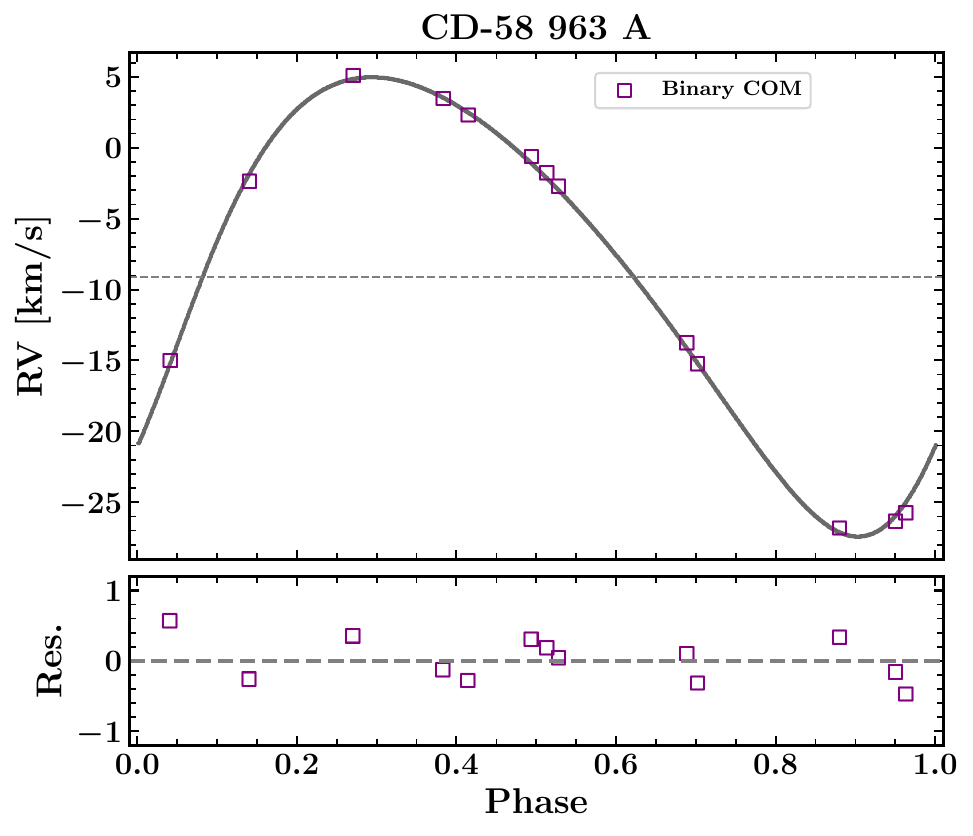}
    \includegraphics[width=0.45\textwidth]{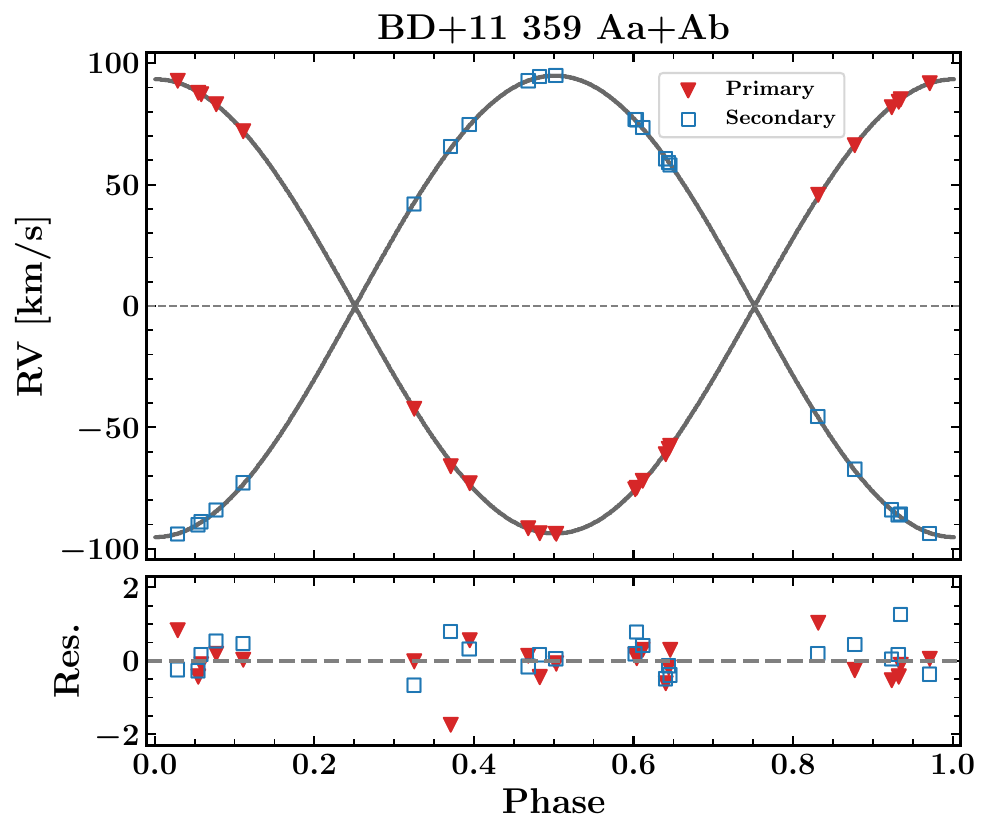}
    \includegraphics[width=0.45\textwidth]{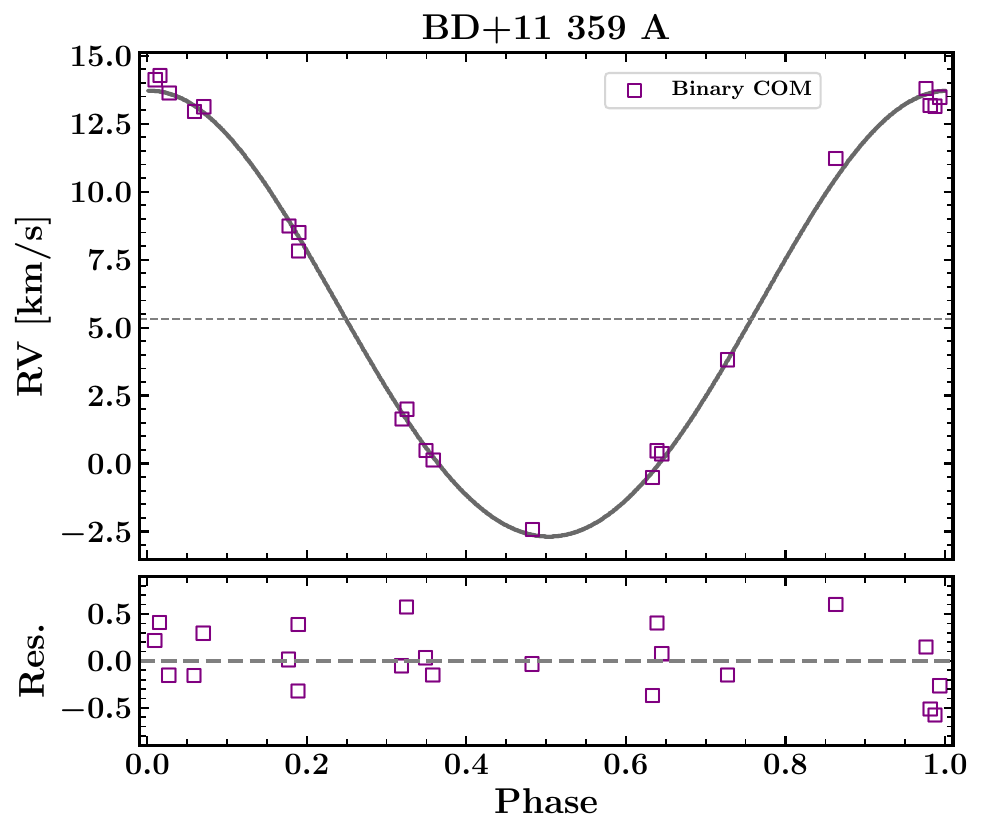}
   
\caption{Same as Fig.\ref{fig:st2rvfit} but for ST2 systems CD58 (top) and BD11 (bottom). Since there are no RVs of the tertiary, the right panels show only the binary's COM.}
 \label{fig:st2rvfit}
\end{figure*}

The results of our orbital RV fits are presented in Table~\ref{tab:v2fit}. The RV measurements and modelled curves are shown in \autoref{fig:st3rvfit} and \autoref{fig:st2rvfit} for the ST3 and ST2 cases, respectively. Individual measurements are given in Table~\ref{appx:rvs}.

\begin{table*}[]
    \centering
     \caption{Orbital parameters of the binary and tertiary orbit obtained by RV modelling.}
    \label{tab:v2fit}
    \begin{tabular}{lcccc}
    \hline
         Parameters &  CD32 & CD62  & CD58 & BD11\\
         \hline
         \hline
         \multicolumn{5}{c}{Binary orbit} \\
         \hline
         $P_A$  [d] &  4.0217043(20) & 2.7147422$^a$ &3.5519749(47) &3.6049188(63)\\ 
         $T_{pA}$ [BJD - 2450000] &  5917.6859(40) & 6952.0938(13)  & 8385.0959(3920) & 2448.3527(64)\\
         $e_A$ & 0.2276(12) & 0$^a$ & 0.00028(22) & 0$^a$\\ 
         $\omega_A$ [deg] & 128.45(30) & 0$^a$ &161.46(39.41) & 0$^a$\\
         $K_{Aa}$ [km/s] & 92.27(26) & 98.75(36) )&86.11(31)&94.90(14)\\
         $K_{Ab}$ [km/s]& 98.97(13) & 109.29(41)&93.86(38)&93.43(12)\\
         $a_{A} \sin{i_\mathrm{A}}$ [$R_{\odot}$] & 14.807(23) &11.166(30)  &12.639(34) &13.423(13)\\
         $ \gamma_{A} $ [km/s] & 1.58(20)  & -23.35(36) &-9.12(13) &5.31(10)\\
         \hline
         \multicolumn{5}{c}{Tertiary orbit} \\
         \hline
         $P_{AB}$ [d]& 1372.1218$^a$ & 441.615(110)  & 76.319(169) &168.581(305)\\
         $T_{pAB}$ [BJD - 2450000] & 5237.44(6.4) & 5866.33(2.79) &5414.62(10.05) &5340.47(15.05)\\
         $e_{AB}$ & 0.2688(50) & 0.2997(144)  &0.2148(96) & 0.0250(209)\\
         $\omega_{AB}$ [deg] & 38.89(1.67) & 217.57(1.91)  & 232.97(3.06) & 359.22(36.28)\\
         $K_{A}$ [km/s] & 6.28(18) & 17.48(96) &16.20(18) & 8.20(15) \\
         $K_{B}$ [km/s] & 17.65(08) & 27.47(64) &- & -\\
         $a_{AB}\sin{i_\mathrm{AB}}$ [$R_{\odot}$] & 625.44(6.610) & 374.49(10.13) &- & -\\
         $ \gamma_{AB}$ [km/s]  & 1.45(7)  & -24.76(23)& -&- \\
         \hline
         \multicolumn{5}{c}{Mass estimates} \\
         \hline
         $M_{Aa} \sin^3{i_\mathrm{A}}$ [$M_{\odot}$] &1.3928(6) &1.3305(11)  & 1.1190(100) &1.2377(36)\\ 
         $M_{Ab} \sin^3{i_\mathrm{A}}$ [$M_{\odot}$]&1.2986(8) & 1.2023(9)  & 1.0266(84) & 1.2573 (40)\\
         $M_{A} \sin^3{i_\mathrm{AB}}$ [$M_{\odot}$]&1.2846(247) & 2.2058(1511)&-&-\\
         $M_{B} \sin^3{i_\mathrm{AB}}$ [$M_{\odot}$] &0.4576(201) & 1.4045(1437)&-&-\\
         \hline
    \end{tabular}
        \vspace{0.2cm}

        $^a$ Fixed during optimisation. \\

\end{table*}

\subsection{Light curve fitting}
We used version 40 of the \textsc{jktebop} code \citep{jktebop} for our LC modelling. \JKTEBOP{} models a star as a sphere or as a biaxial spheroid and calculates the LC by numerical integration of concentric circles. This allows it to fit only detached EBs. With binary periods of more than two days, \JKTEBOP{} is well suited for solving our systems. We modelled every TESS{} sector separately except for CD58, where we modelled only half of the LC for a sector. In our modelling process, we first fixed certain parameters from RV modelling, and/or from prior knowledge about the type of stars. The fixed parameters are (i) mass-ratio ($q$), and (ii) limb-darkening coefficients. We took initial values for \Pin{} from the  RV solution but varied them during our modelling. We also took a visual estimate of the time of super-conjugation ($T^\mathrm{A}_\mathrm{0}$) and optimised it later. We optimised the following LC parameters: (i) \Pin, (ii) \Toin, (iii) the scale-factor (which determines the scaling or the magnitude of the out-of-eclipse portion), (iv) the surface brightness ratio ($J$), (v) the third light ($L_3$), (vi) \ein{} and $\omega$ in the form of $e\sin{\omega}$ and $e\cos{\omega}$, (vii) the inclination of inner binary (\incin), (viii) the radius ratio ($k$), and (ix) the sum of fractional radii ($r_1 + r_2$), where the fractional radii are represented as the radius divided by the semi-major axis (\smain). The optimisation was iterated until we got the best fit. To test the convergence, we randomly fixed certain parameters and optimised the others to check the stability. We then estimated the errors on the parameters by using the Monte Carlo (MC) module available on \JKTEBOP. We followed this exact prescription for all the targets except CD58. CD58 has shallow eclipses (the primary eclipse has a depth of $\sim$0.04 mag) and shows stellar activity that varies with every sector. On top of that, the tertiary orbital period is 77 d. This causes eclipse timing variations (ETVs) that affect radii measurement in one TESS{} sector and that is why we used half a sector for our modelling. To get a consistent solution over sectors, we first fixed the third light to zero. \JKTEBOP{} allows one to set the light ratio from spectroscopy as a constraint, so we constrained the optimisation by fixing the light ratio of the components as the values obtained from broadening function (BF) fitting (see \autoref{sec:BF}). The rest of the optimisation process was the same as before. The MC sampling for CD58 was initiated with the final optimised parameters but without constraints from the light ratio. The final fits to the LC are shown in \autoref{fig:LC}. The final results for all the targets for all the modelled sectors are given in Appendix.\ref{appx:jktebop}. We adopted the final estimates of all the parameters as the average of all the sectors. \\

\subsection{Estimate of orbital configuration}
Combining both the LC and RV analyses, we can get a picture of the orbital architecture of the systems, except the estimate of the mutual inclination (\incmut). This is possible for both ST2 and ST3 systems if we observe dynamical effects in ETVs. For ST3 systems specifically, we can estimate \incmut{} using inclinations of the binary (\incin) and tertiary (\incout) orbits. Using equations described in \cite{moharanacht}, we get two different sets of limits for the \incmut{}. With the estimated value of \incmut{} for ST3 systems, and an arbitrary value for ST2 systems, we can simulate the orbits using numerical integration. We used \rebound\footnote{\url{https://github.com/hannorein/rebound}} \citep{rebound}, an n-body numerical integration code, to obtain the architecture of the systems. The configuration of the orbits in the XZ plane (the ~Z axis is towards the observer) is shown in \autoref{fig:orbit}.

\begin{figure*}
    \centering
        \includegraphics[width=\textwidth]{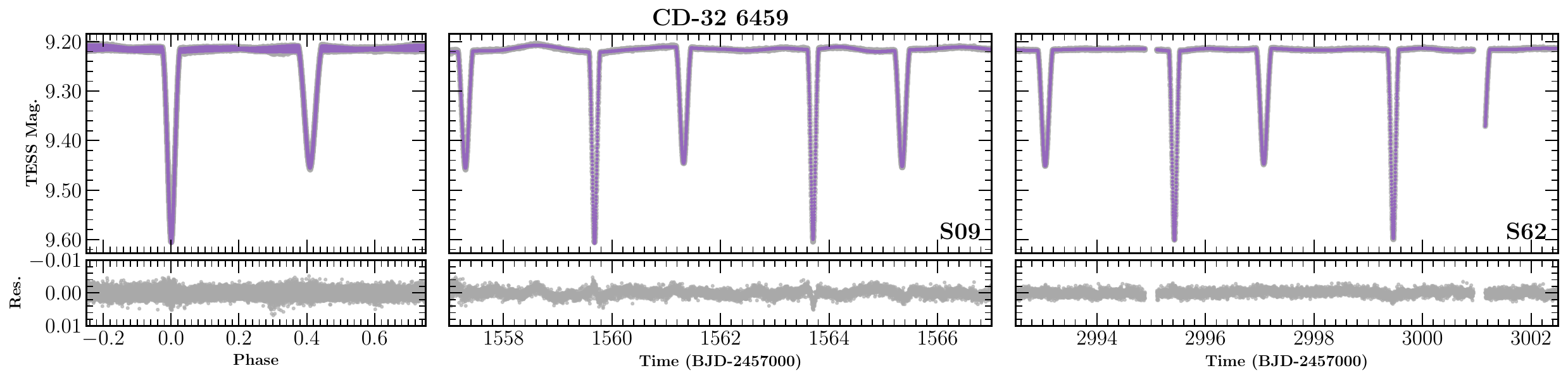}
        \includegraphics[width=\textwidth]{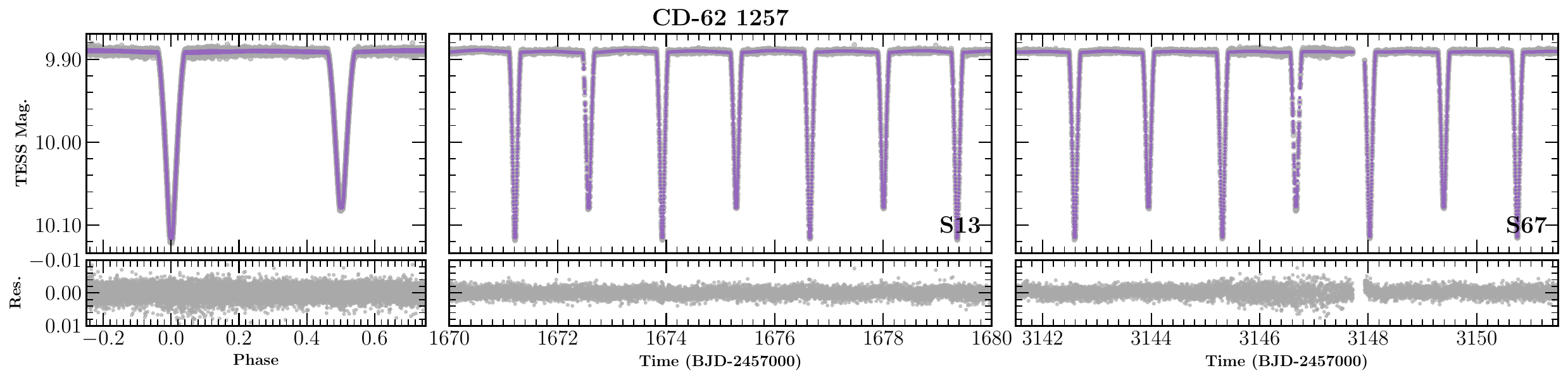}
        \includegraphics[width=\textwidth]{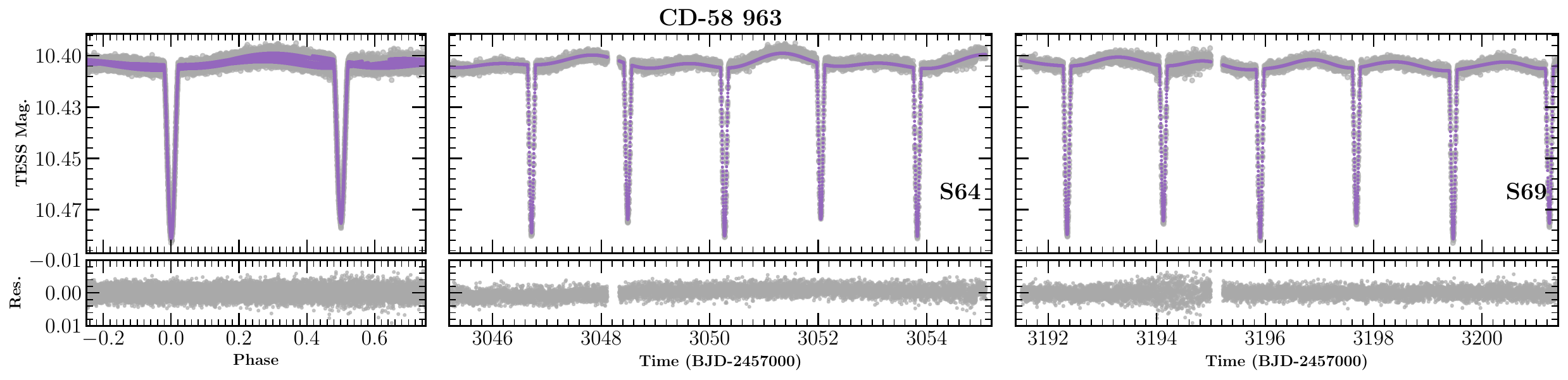}
        \includegraphics[width=\textwidth]{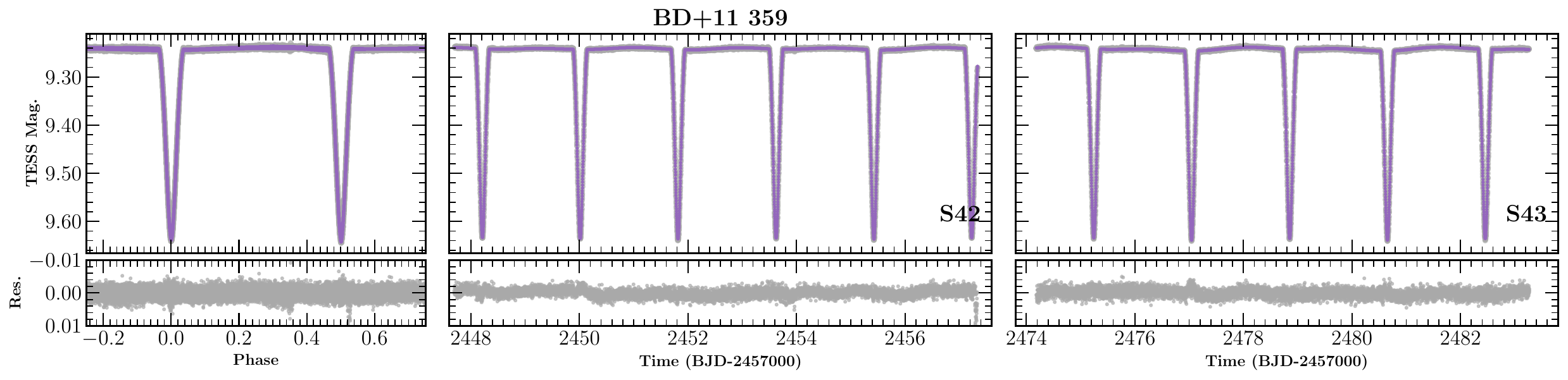}
    \caption{\JKTEBOP{} models (purple lines) for the first and the last used in our modelling of TESS{} observations (centre and last panel, respectively). The variable stellar activity can be seen in the sector-wise observations and models. The eccentric and near-circular systems can be distinguished from the phased LC observations and models (first panel). }
    \label{fig:LC}
\end{figure*}

\begin{figure*}
    \centering
    \includegraphics[width=0.45\textwidth]{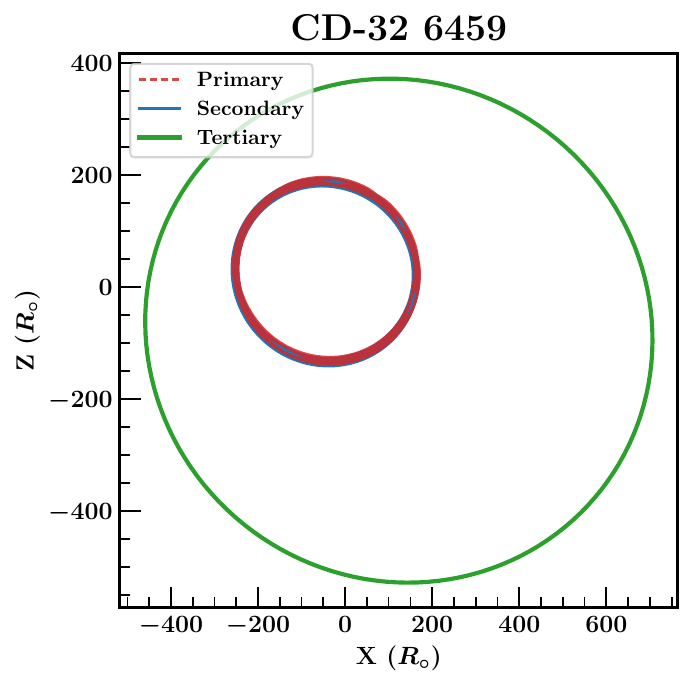}
     \includegraphics[width=0.45\textwidth]{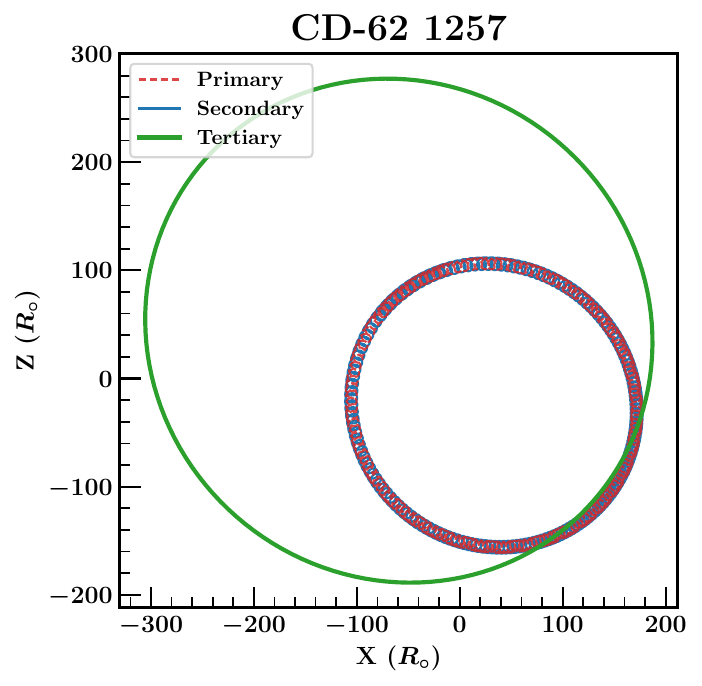}
       \includegraphics[width=0.45\textwidth]{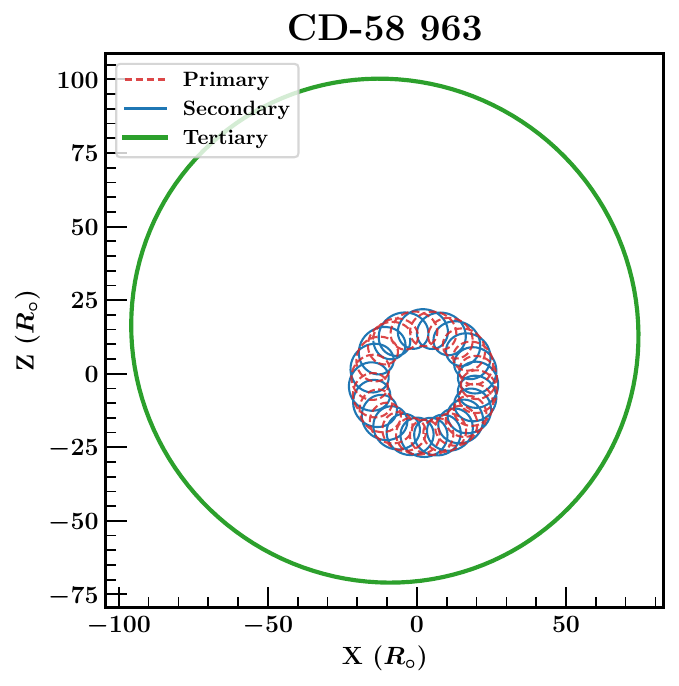}
      \includegraphics[width=0.45\textwidth]{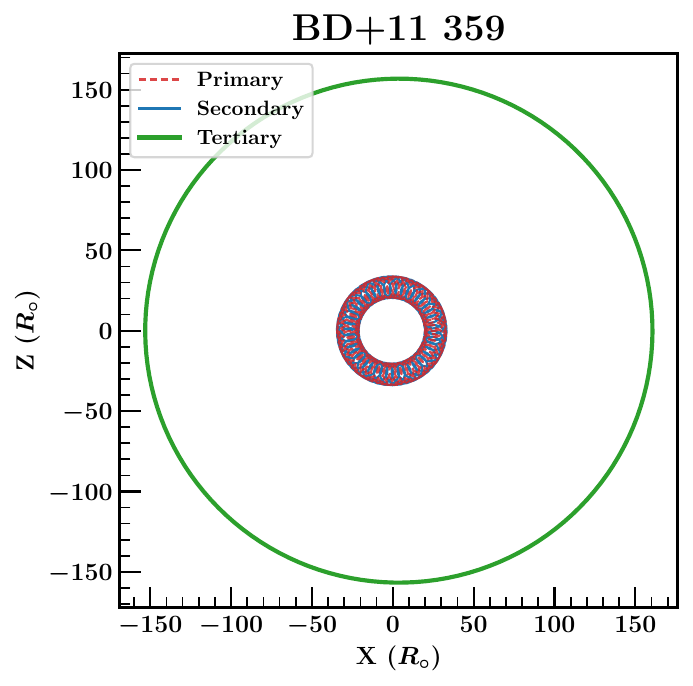}
    \caption{Orbital configuration of the targets viewed perpendicular to the observers' plane (XY plane). The COM of the system is at (0,0).}
    \label{fig:orbit}
\end{figure*}

\subsection{Spectroscopic analysis}
\label{sec:specanal}
For our spectroscopic analysis, we need a set of homogeneous spectroscopic data; in other words, from a single instrument. We selected a set of time-series spectroscopy for each target with clean line profiles and good S/N. This resulted in a set of 9 CHIRON spectra for CD32, 7 HRS spectra for CD62, 13 HRS spectra for CD58, and 8 FEROS spectra for BD11. All further analyses were done on these sets of spectra.

\subsubsection{Broadening functions}
\label{sec:BF}
We used BFs (\citealt{bfsvd}) to (i) estimate light fractions ($LF^{x}_\mathrm{obs}$), and (ii) estimate the projection of rotational velocity (\vsin{x}) of the component, $x$. The BF was generated with a variation of the BF code, \textsc{bf-rvplotter}\footnote{\url{https://github.com/mrawls/BF-rvplotter}}. Clean BFs (where we can see all spectral components) were modelled with the rotational profile from \citet{stellarphotgray} using the formula
\begin{equation}
    G(v)=A \left [c_{1}\sqrt{1-\left(\frac{v}{v_\mathrm{max}}\right)^{2}}+c_{2}\left (1-\left (\frac{v}{v_\mathrm{max}}\right)^{2}\right )\right] +l v+k ,
\end{equation}
    where $A$ is the area under the profile and $v_\mathrm{max}$ is the maximum velocity shift that occurs at the equator. The terms $c_{1}$ and $c_{2}$ are constants that are a function of limb darkening themselves, while $l$ and $k$ are correction factors to the BF `continuum'.  The final $LF^{x}_\mathrm{obs}$, and \vsin{x} values were taken as the average of all the epochs of spectra, with respective standard deviations as errors. These values are presented in the lower panel of \autoref{tab:spec_anal}.

\subsubsection{Spectral disentangling}
 We separated the individual spectra from the composite spectra using the method of spectral disentangling with the shift-and-add algorithm \citep{shiftandadd} implemented in the code \textsc{disentangling\_shift\_and\_add}\footnote{\url{https://github.com/TomerShenar/Disentangling_Shift_And_Add}} (\textsc{dsaa}; \citealt{shenardsaa2020,shenardsaa2022}). \textsc{dsaa} disentangles spectra in the velocity space by applying simple velocity shifts corresponding to one component and creating an averaged spectrum. This is iterated for every component until the final individual spectra have no contribution from the rest of the components. The code takes in orbital parameters along with a list of times of epoch to disentangle the spectra.
 We disentangled spectra in the wavelength range from 4870 \AA{} to 5300 \AA . This region was selected to avoid wide lines and due to the availability of sufficient narrow lines for spectral analysis. We initially assumed equal light fractions for the component. The final disentangled spectra were accepted after using \textsc{dsaa} for 30,000 iterations.  In addition to the convergence checks built into the code, we also calculated cross-correlation plots on the final output to check for any contaminant lines in the spectra due to the disentangling process. The disentangled spectra had trends in the continuum, which were a result of bias in normalisation, light-fraction variation, et cetera \citep{spuriouspatt}. The amplitude of these trends varies depending on the number, the extent of convolution of line profiles, and the wavelength range of the spectra used for disentangling. Further, it depended on the spectrograph, which we attribute to either the stability of the spectrograph or the accuracy of the spectral reduction method. To address the first source of bias, we selected the best spectra by preliminary trials with the wavelength range and set of spectra. Further, we cleaned this additive form of bias by modelling this trend and subtracting it from the component spectra, following the process in \cite{spuriouspatt}. 
 We then scaled the spectra for a component, $x$, to the observed values (from BF) of the light fraction, using the formula
 \begin{equation}
     f^{x}_\mathrm{new}=(f^{x}_\mathrm{ini}-1)\times (LF^{x}_\mathrm{ini}/LF^{x}_\mathrm{obs})+ 1
 .\end{equation}

\subsubsection{Synthetic spectral fitting}

\begin{figure*}
    \centering
    \includegraphics[width=\textwidth]{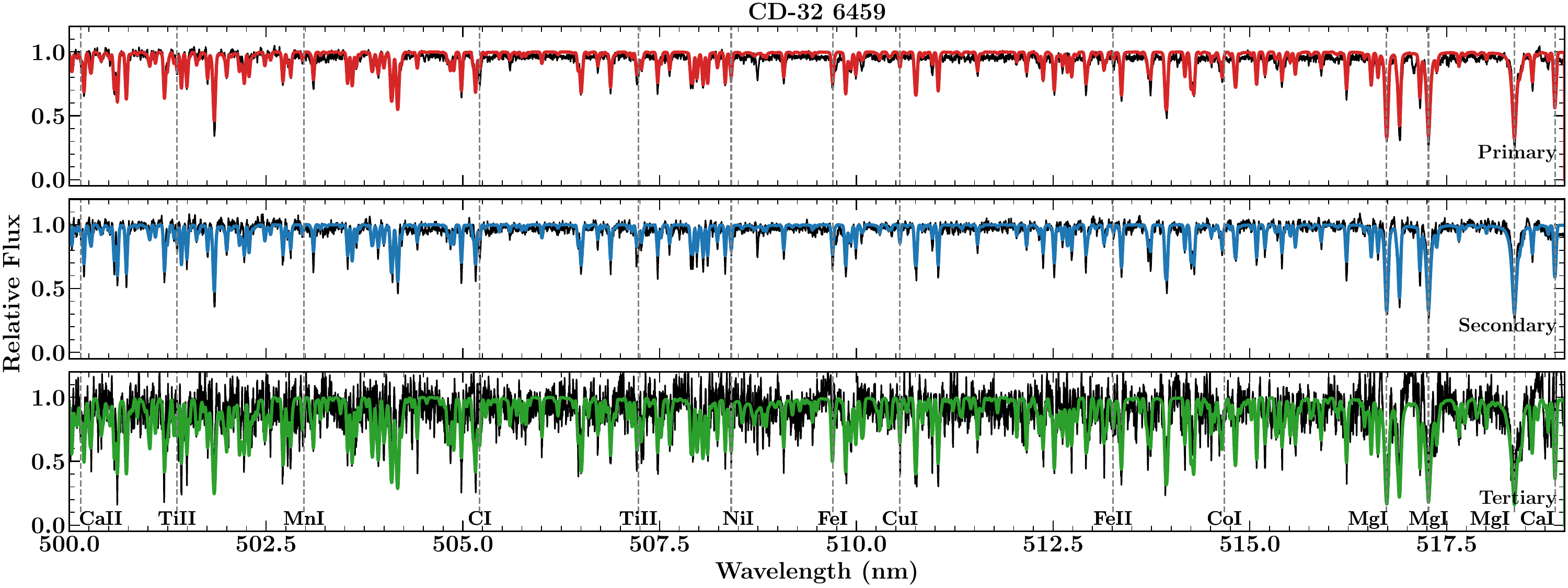}
    \includegraphics[width=\textwidth]{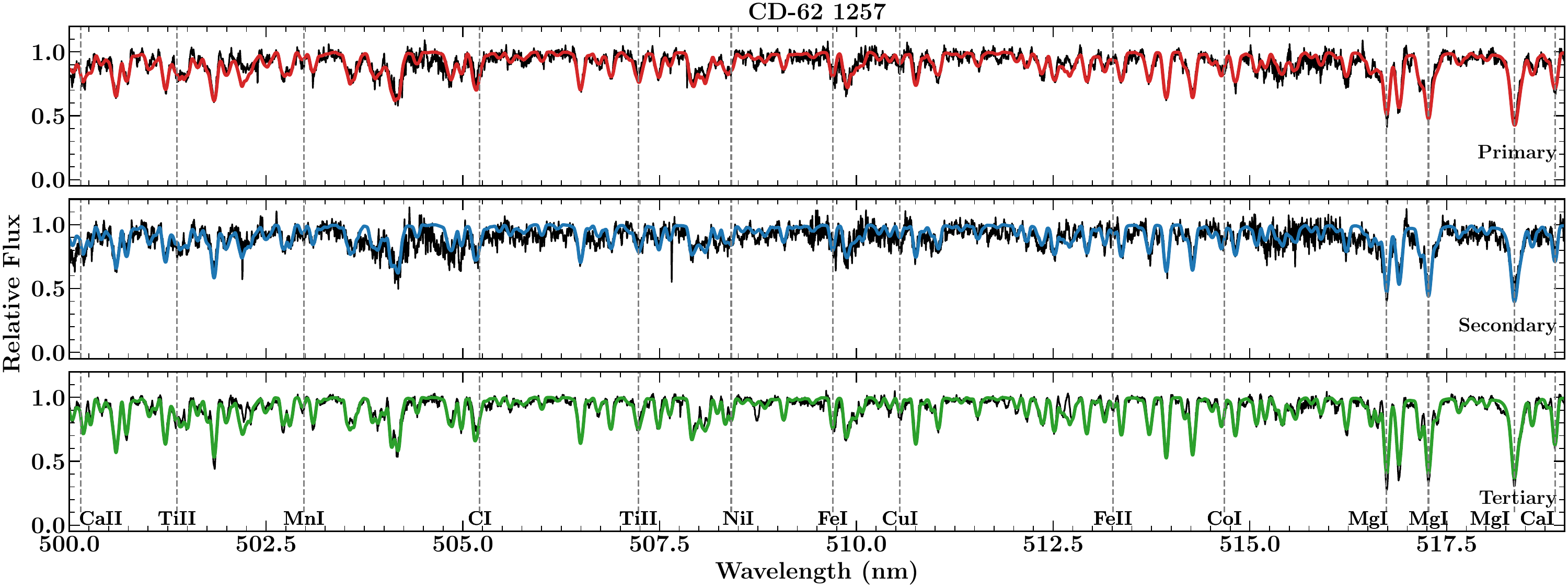}
    \includegraphics[width=\textwidth]{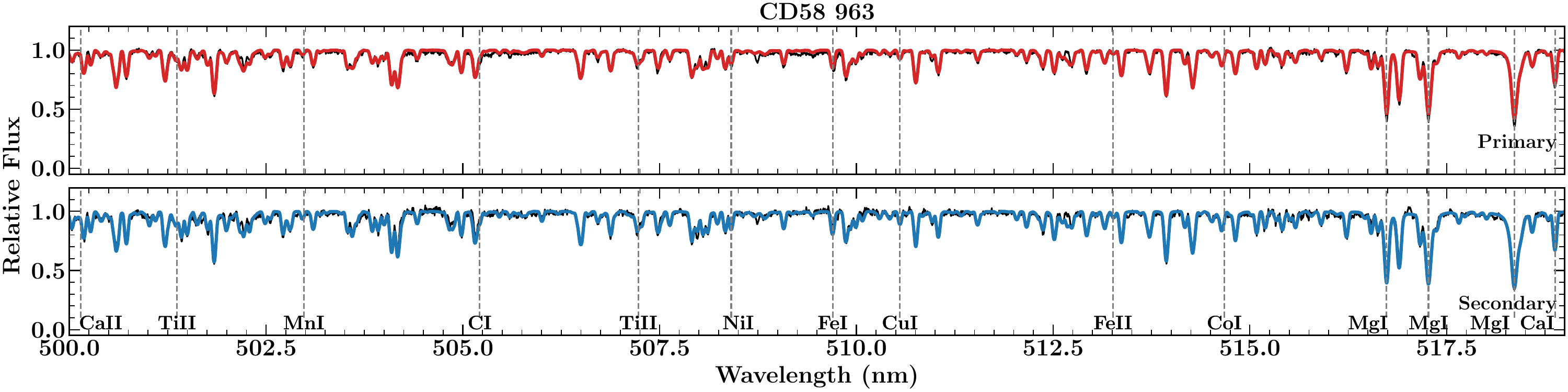}
    \includegraphics[width=\textwidth]{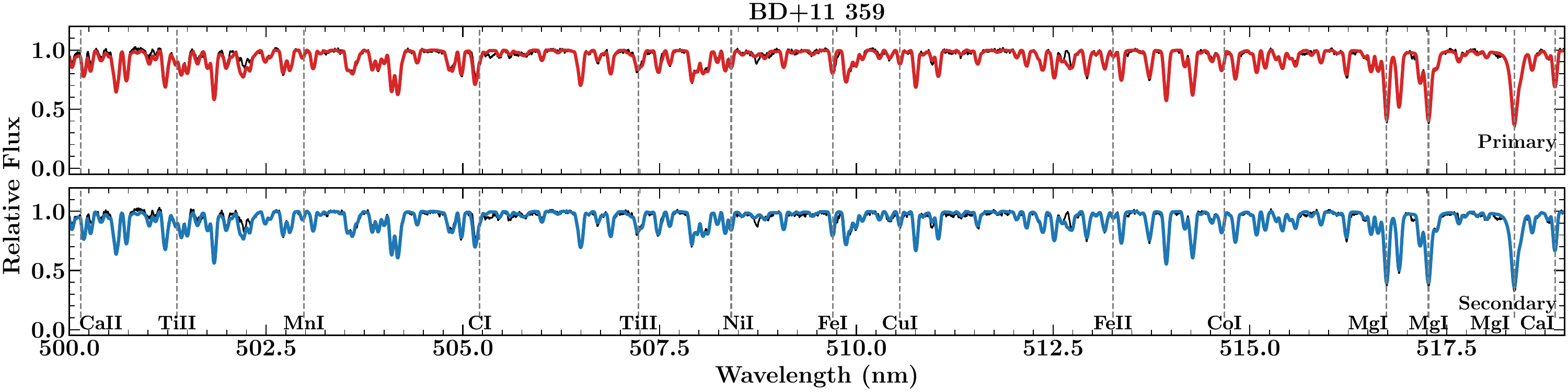}
    \caption{Synthetic spectroscopic models of primary (red), secondary (blue), and tertiary (green) spectra compared to the disentangled spectra (black) for the different systems. The panels for ST2 systems (CD58 and BD11) show only the primary and secondary spectral models.}
    \label{fig:spec_all}
\end{figure*}

 For the measurement of effective temperature ($T_\mathrm{eff}$), metallicity ($[M/H]$), and log of surface gravity ($\log{g}$), we used \textsc{ispec} \citep{ispec2014,ispec2019} on the disentangled spectra. All the disentangled spectra were checked for any RV offset caused by the disentangling method. We did not normalise the spectra further, as the \spd{} and the bias-cleaning already give us normalised spectra. We obtained the atmospheric parameters using the synthetic spectral fitting (SSF) technique. The SSF technique generates synthetic spectra on the go and then does a $\chi^{2}$ optimisation at selected spectral lines. This method is better than a simple grid fitting \citep{ispec2019}. We implemented different fitting procedures for the eclipsing stars and the tertiary, respectively. For the eclipsing stars, we fixed the \logg{} estimated from LC and RV modelling using
 \begin{equation}
\log (g)=\log \left( \frac{A^2_c M}{R^2} \right),
\end{equation}
where $M$ is the mass in units of M$_\odot$, $R$ is the radius in R$_\odot$, calculated from LC and RV modelling, and $A_c \equiv \sqrt{G M_\odot}/R_\odot (= 168.589888477)$ is a constant necessary for transformation to solar units. The resolution of FEROS and HRS was high enough to give us the precise projected rotational velocity ($v\sin{i}$) from SSF, but for CHIRON we fixed it to the value obtained with BF. 
The set-up for synthetic spectra generation includes model atmospheres from MARCS\footnote{\url{https://marcs.astro.uu.se/}} \citep{marcs}, solar abundances from \cite{asplund2009}, and the radiative transfer code \textsc{spectrum}.\footnote{\url{https://www.appstate.edu/~grayro/spectrum/spectrum.html}} We calculated the parameters using two different line lists. We first fitted for $T_\mathrm{eff}$, $[M/H]$, $\alpha$, and $v\sin{i}$ using line list LL1, which is prescribed for abundance measurement. We adopted $[M/H]$ and $\alpha$ from this run and then fitted for $T_\mathrm{eff}$ and $v\sin{i}$ using line list LL2, which is prescribed for parameter estimation. For the eclipsing systems, we kept the $\log{g}$ fixed as the values (\autoref{tab:paramtablest2} and \autoref{tab:paramtablest3}) that we obtained from LC and RV modelling, since the spectroscopic $\log{g}$ matched well but had a lower precision. 
We kept the $\log{g}$ free for the tertiary spectra. The final spectroscopic estimates for all the systems are given in \autoref{tab:spec_anal} and the best-fit models are shown in \autoref{fig:spec_all}.

\begin{table}[]
    \centering
    \small
     \caption{Atmospheric parameters obtained from our spectroscopic analysis on the disentangled spectra.}
    \label{tab:spec_anal}
   \begin{tabular}{lcccc}
    \hline
         Parameters &  CD32 & CD62 & CD58 & BD11\\
         \hline
         \hline
        \multicolumn{5}{c}{Synthetic spectral fitting} \\
         \hline
$T_\mathrm{eff,Aa}$ & 6674(248) & 6729(263) & 6478(171) & 6342(181) \\
$T_\mathrm{eff,Ab}$ & 6295(244) & 6525(262) & 63712 (177) & 6553(178) \\
$T_\mathrm{eff,B}$ & 6344(412) & 6623(143) & - & -\\
$\log{g}_\mathrm{Aa}$ & 4.21$^a$ & 4.17$^a$ & 4.25$^a$ & 4.14$^a$ \\
$\log{g}_\mathrm{Ab}$ & 4.25$^a$ & 4.23$^a$ & 4.35$^a$ & 4.17$^a$ \\
$\log{g}_\mathrm{B}$ & 3.8(5) & 4.1(1) & - & -\\
$[M/H]_\mathrm{Aa}$  & -0.24(5) & 0.31(7) & -0.34(6) & -0.07(2) \\
$[M/H]_\mathrm{Ab}$ & -0.26(1) & 0.37(9) & -0.28(4) & -0.16(1) \\
$[M/H]_\mathrm{B}$ & 0.66(3) & 0.13(7) & - & -\\
$\alpha_\mathrm{Aa}$ & 0.12(6) & -0.00(8) & 0.05(6) & -0.13(3) \\
$\alpha_\mathrm{Ab}$ & -0.28(1) & -0.05(9) & 0.08(4) & 0.08(2) \\
$\alpha_\mathrm{B}$  & 0$^a$ & -0.00(5) & - & -\\
$v_\mathrm{mic,Aa}$ & 1.5(3) & 1.5(3) & 1.7(3) & 2.1(1) \\
$v_\mathrm{mic,Ab}$ & 0.1(1) & 0.0 & 1.4(2) & 2.6(1) \\
$v_\mathrm{mic,B}$ & 1.49$^b$ & 2.4(2) & - & -\\
$v_\mathrm{mac,Aa}$  & 11.06$^b$ & 11.05$^b$ & 9.25$^b$ & 7.8$^b$ \\
$v_\mathrm{mac,Ab}$ & 7.59$^b$ & 8.88$^b$ & 9.32$^b$ & 9.76$^b$ \\
$v_\mathrm{mac,B}$ & 8.25$^b$ & 15.38$^b$ & - & -\\
$v_\mathrm{Aa}\sin{i_A}$ & 0$^a$ & 30(2) & 20(1) & 23.90(39) \\
$v_\mathrm{Ab}\sin{i_A}$ & 0$^a$ & 31(2) & 18(1) & 21.54(43) \\
$v_\mathrm{B}\sin{i_{AB}}$ & 0$^a$ & 18(1) & - & -\\
             \hline
         \multicolumn{5}{c}{Broadening functions}\\
         \hline
         $v_\mathrm{Aa}\sin{i_A}$ & 15(2) & 33(3)  &23(3) &27.06(51)\\ 
         $v_\mathrm{Ab}\sin{i_A}$ &  11(3) &27(2) & 19(3) & 25.99(40)\\
         $v_\mathrm{B}\sin{i_{AB}}$ &  4(4) &25.5(5)  & - & -\\
         \hline
         \vspace{0.1cm}
         $LF^{Aa}_\mathrm{obs}$ &  0.526(3) & 0.28(1)  & 0.60(2) & 0.526(2)\\
         \vspace{0.1cm}
         $LF^{Ab}_\mathrm{obs}$ &  0.375(5) & 0.16(2)  & 0.40(2) & 0.474(2)\\
         $LF^{B}_\mathrm{obs}$ &  0.099(4) & 0.56(2) & - & -\\

    \hline
    \end{tabular}
    
        $^a$ Fixed during optimisation. $^b$ Obtained from empirical tables.  
\end{table}

\subsection{Isochrone fitting}
Most CHTs with detailed analysis have been found to have co-evolving stars. This implies that we can use a single isochrone to explain the masses, radii, and temperatures of all stars. While there have been cases in which non-coeval stars seem to exist in a CHT \citep{hd144noncoeval,fredv12}, we used this assumption to check and estimate the age of the system as a whole.  We used our fitting code \textsc{isofitter}, which searches through the grid of MIST (MESA Isochrones and Stellar Tracks; \citealt{mistc2016,mistd2016}) and finds the best-fit model that explains the masses, radii, metallicities, and temperatures of all three stars. There is also an option to add \Gaia{} distances and flux ratios as constraints in the minimisation routine. Details of the routine can be found in \cite{moharanacht}. We made two runs for the ST3 systems. In one run, we used the observed constraints from all three stars, and in the other we used only the inner binary (also adding the binary flux ratio as a constraint). The best-fit tertiary models (in black) and binary models (in purple) are shown in \autoref{fig:cd32isofit} and \autoref{fig:cd62isofit} for CD32 and CD62, respectively. The same routine was used for the ST2s, with their stellar parameters, flux ratio, and {\it Gaia} distance used as constraints. Assuming co-evolution, we can also estimate the third star's radius for a range of tertiary masses. We used this to estimate the radius and temperature of the ST2 tertiaries, assuming $M_3$ between the lower limit from RV fits and $M_2$. For ST2s, we also rejected models that predict a distance of more than 3$\sigma$ from the {\it Gaia} value (\autoref{appx:isofitter}).

\section{Results and discussion}
In the following text, we use A-B notation to denote the CHT, where B is the tertiary and A is the EB with components Aa and Ab. Here, Aa corresponds to the primary, which is the most massive star in the inner binary. We use the short form for a star's name along with the alphabetical notation to exclusively denote each star; for example, the secondary of CD62 is referred to as CD62Ab.
\label{sec:results}
\subsection{CD-32 6459 $\equiv$ TIC024972851}
CD32 is the widest of all our systems (largest $a_{AB}$ value). The inner binary is also the largest of all four systems (largest $a_{A}$ value). The binary masses are above-solar with a primary and secondary mass of 1.406 $M_{\odot}$ and 1.211 $M_{\odot}$, respectively. With radii of 1.54 $R_{\odot}$ and 1.44 $R_{\odot}$ for CD32Aa and CD32Ab, respectively, both are in the main-sequence regime. It is the only system with a non-negligible eccentricity of $e_A=0.221$, but with an inner binary period of $\sim$4~d, the orbit should be circularised after $\sim$500~Myr \citep[according to the formalism of ][]{zahn1989_2}. The mutual inclination regimes lie in two possible sets of solutions: (i) 34.3$^\circ$ to 136.7$^\circ$ or (ii) 43.3$^\circ$ to 145.7$^\circ$. These limits are close to the limits for Von Zeipel-Lidov-Kozai (ZLK; \citealt{zeipel,lidov,kozai}) oscillations, which could be the reason behind the large value of $e_A$. ZLK oscillations in such close binaries make it a good candidate to see Kozai cycles with tidal friction (KCTF; \citealt{kisleva1998tidalfrict}) in action. Simulations predict that the product of KCTF will have a large period ratio, a mutual inclination close to the critical limit, and a circular inner orbit \citep{fabrytremaine2007}. CD32 shows all the above properties except a circular inner orbit. This makes CD32 a case with which to study ongoing KCTF or a candidate for a failed KCTF process (see estimated age in the following discussions).

\begin{figure}
    \centering
    \includegraphics[width=\columnwidth]{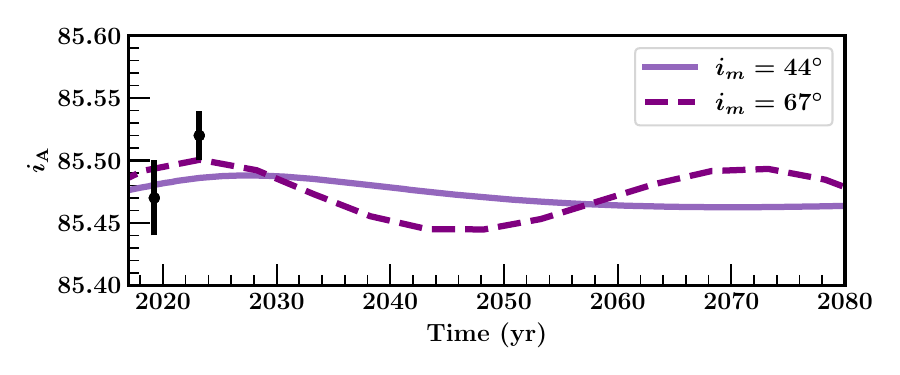}
    \caption{\rebound{} simulations of \incin{} variations (upper panel) of CD32 (in purple). Estimated inclinations from LC modelling are shown with black dots.}
    \label{fig:CD32EDV}
\end{figure}

\begin{figure}
    \centering
    \includegraphics[width=\columnwidth]{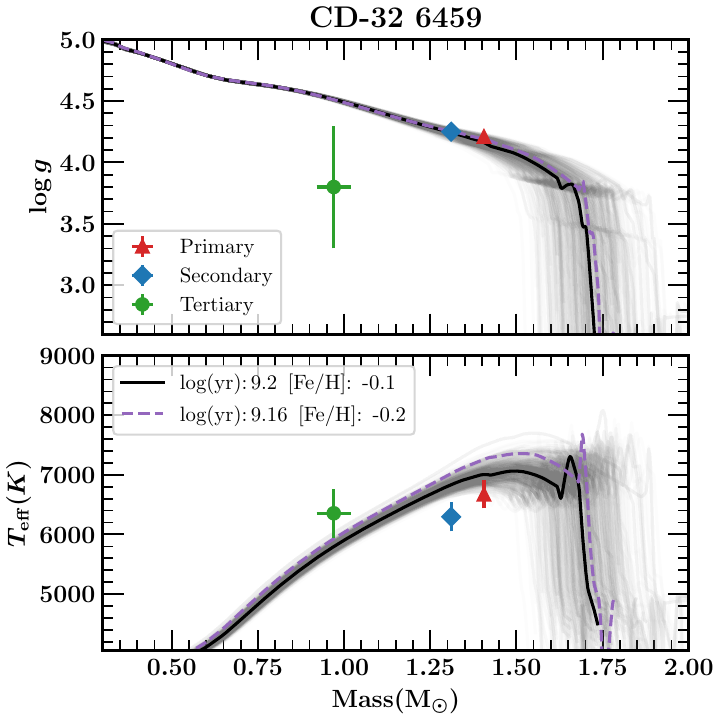}
    \caption{Results of the isochrone fitting for CD32. The black line is the best-fitting isochrone for the run with all constraints (grey lines are exemplary of other isochrones within the constraints), and the purple line is for the run in which only the inner binary was used. Apart from the low $\log{g}$ of the tertiary (mostly due to a noisy disentangled spectrum), the parameters observed are consistent with isochrone estimates within $5\sigma$ errors.}
    \label{fig:cd32isofit}
\end{figure}

\begin{figure}
    \centering
    \includegraphics[width=\columnwidth]{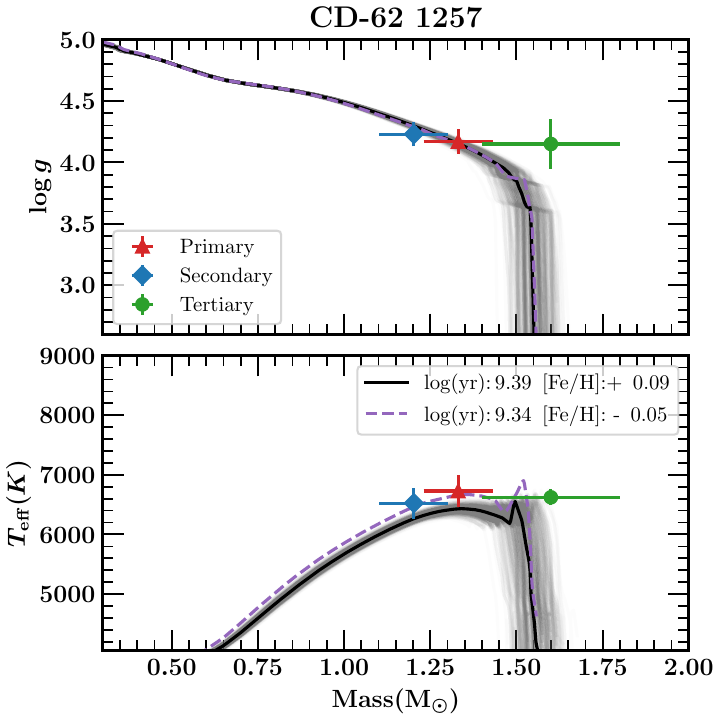}
    \caption{Same as \autoref{fig:cd32isofit} bur for CD62.}
    \label{fig:cd62isofit}
\end{figure}

The ZLK oscillations in CD32 can also cause it to show EDVs. To check the possibility of observing the variations in the future, we simulated the system for a period of 100 years using \rebound{} with \incmut{} values of 44$^{\circ}$ and 67$^{\circ}$. We find that while the cycles themselves have a period of more than 4000 years, the short-term higher-order perturbations have amplitudes close to the precision limits (from LC modelling) for \incin{} values (\autoref{fig:CD32EDV}). This is because according to \cite{triplepertperiods}, the amplitudes for medium-period and short-period perturbations are of the order of $P_\mathrm{A}/P_\mathrm{AB}$ and $(P_\mathrm{A}/P_\mathrm{AB})^2$, respectively, which are small for CD32.  The simulations did not include tidal effects that would further subdue the chances of detecting EDVs.

The disentangled spectrum for CD32B (\autoref{fig:spec_all}; third panel from top) is quite poor because of the faintness of the component, the comparatively low resolution ($\sim$28,000), and a low number of spectra. Therefore, we fixed most of the atmospheric parameters using approximations for metallicity and $\alpha$ from the inner binary. The final \logg{} and radius estimates had large error bars. The temperatures are above solar values and the system is metal-poor (-0.25 dex). Isochrone fitting with these values suggested a log(age) of 9.2 or 1.58 Gyrs, \ie{} more than the inner orbit circularisation timescale. While the estimated $\log(g)$ of CD32B is way off the isochrone value, the temperature matched the isochrone temperature within 5$\sigma$ errors. Because of the poor spectroscopic estimate, we adopted the isochrone radius estimate for the tertiary (0.9 $R_{\odot}$). Isochrones from binary fitting were consistent with the tertiary one. Therefore, we consider this system to be a co-evolving system. 

\subsection{CD-62 1257$\equiv$ TIC387107961}
\begin{figure}
    \centering
    \includegraphics[width=0.9\columnwidth]{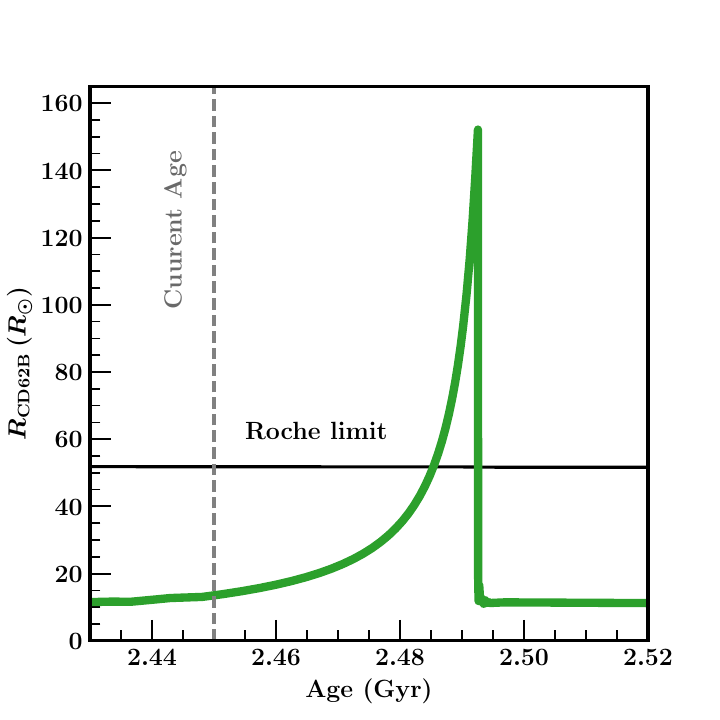}
    \caption{Radius evolution of CD62B. The star evolves in the outer orbit of the CHT and reaches the Roche limit after 40 Myrs. This can prompt a mass transfer that leads to a TCE system.}
    \label{fig:CD62TCEE}
\end{figure}

CD62 is the only system in the sample that has a tertiary more massive than the inner binary stars ($M_3=1.6 M_{\odot}$). This is visible in the high third light in the LC and also prominent in the spectra and the BFs. The binary is relatively close, with a period of 2.71456 d and a semi-major axis of 11.17 $R_{\odot}$. The outer orbit is eccentric ($e_{AB}\sim0.3$). The mutual inclination ranges are (i) 16.4$^\circ$ to 162.4$^\circ$ or (ii) 17.6$^\circ$ to  163.6$^\circ$. Spectroscopic analysis shows that the system is metal-rich, with the inner binary stars showing an [M/H] value of 0.3 dex and the tertiary being at 0.13 dex. Isochrone fitting gives us an age of 2.45 Gyr for the tertiary-constrained case and 2.19 Gyr for the binary-constrained case but with a negative metallicity. The more massive tertiary seems to be still on the main sequence even though the isochrone \logg{} expects it to be in the sub-giant phase. While the spectroscopic estimations are precise (compared to CD32B), our fitting is affected by large errors in the CD62B's mass. While it is consistent with co-evolving temperatures, we do not completely rule out the case that the tertiary is evolving differently with its different metallicity and log(g). With the current configuration, CD62B seems to be evolving towards the red giant branch (RGB). Since the tertiary mass ratio is the highest in the sample, we checked for the possibility of Roche lobe mass transfer and subsequently a triple common envelope (TCE) formation. We used a MESA evolutionary grid for a 1.6 $M_{\odot}$ star to see the radius evolution. The calculation of the Roche limit was done using the expression in  \cite{eggletonrocheradii} and the assumption that the CHT is a wide binary with the eclipsing pair as a single star with a mass of $\sim$ 2.5 $M_{\odot}$. We found that the tertiary will evolve beyond the Roche limit in 20 Myr and could form a TCE (see \autoref{fig:CD62TCEE}).

\subsection{CD-58 963 $\equiv$  TIC220397947}
CD58 has the shortest outer orbital period in our sample and (most likely) the smallest $a_{AB}$ value. \cite{Borko2020} found a solution in which the two stars, despite having slightly different masses (1.15 and 1.10~M$_\odot$ at $i=82.3^\circ$), had very similar radii ($\sim$1.21~R$_\odot$). This led to the conclusion that the system is young (18.2~Myr). In our solution, the radii are significantly different from each other, and the inclination is somewhat lower. We obtained a good isochrone fit to our results for the age of 3.01~Gyr, with all the parameters (including {\it Gaia} distance) properly reproduced. It places all three stars on the main sequence. Despite having masses close to the solar one, the Aa and Ab components are significantly hotter, due to their lower metallicity. 

\begin{figure}
    \centering
    \includegraphics[width=\columnwidth]{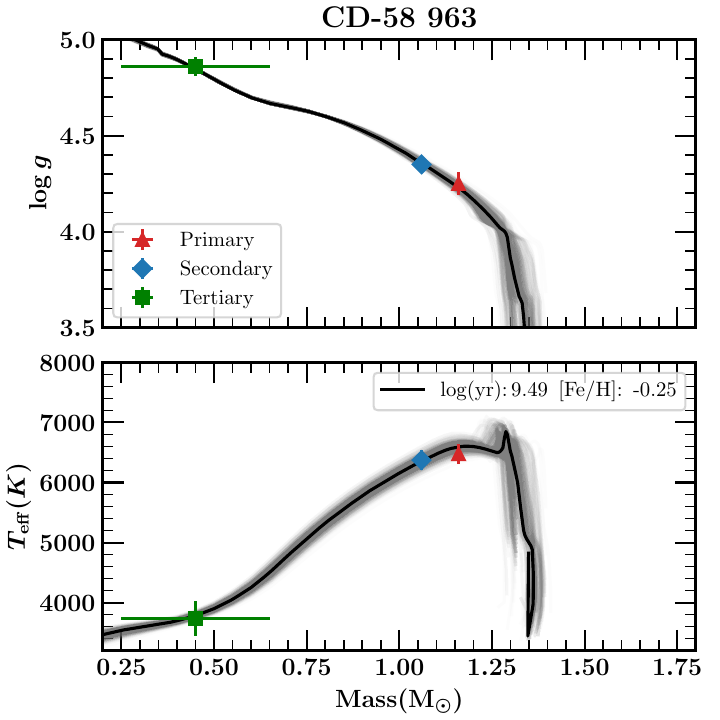}
    \caption{Results of the isochrone fitting for CD58. The parameters observed for Aa+Ab are consistent with isochrone estimates within $3\sigma$ errors. Tertiary parameters are evaluated
    from the isochrone fit.}
    \label{fig:CD58isofit}
\end{figure}

From the isochrone fit (\autoref{fig:CD58isofit}), we can also estimate the properties of the tertiary: $M_B=0.45$~M$_\odot$, $R_B=0.42$~R$_\odot$, and $T_{\mathrm{eff},B}=3730$~K. The value of $M_B$ is close to the lower limit from the RV solution, suggesting a nearly edge-on outer orbit, and likely a co-planar geometry.

There are several discrepancies between our results and \cite{Borko2020}, mainly in the radii of Aa and Ab, and the properties of the tertiary. One of the reasons behind those discrepancies might be the extent of TESS{} data. New observations were obtained during Cycle 5 \citep[][ only had access to Cycle 1]{Borko2020}. By that time, the shape of the LC has changed (see Fig.~\ref{fig:LC}), leading to different estimates of the radii and inclination. Furthermore, \citet{Borko2020} did not have any RV measurements, only ETVs of the inner binary, and thus no direct dynamical reference scale for masses and orbit sizes. We also argue that the presence of the fourth (0.21~M$_\odot$) body at the 2661d orbit did not affect our analysis, since its contribution to the total flux seems to be negligible, and our RV data were taken within a much shorter time frame.

\subsection{BD+11 359 $\equiv$  TIC408834852}
\begin{figure}
    \centering
    \includegraphics[width=\columnwidth]{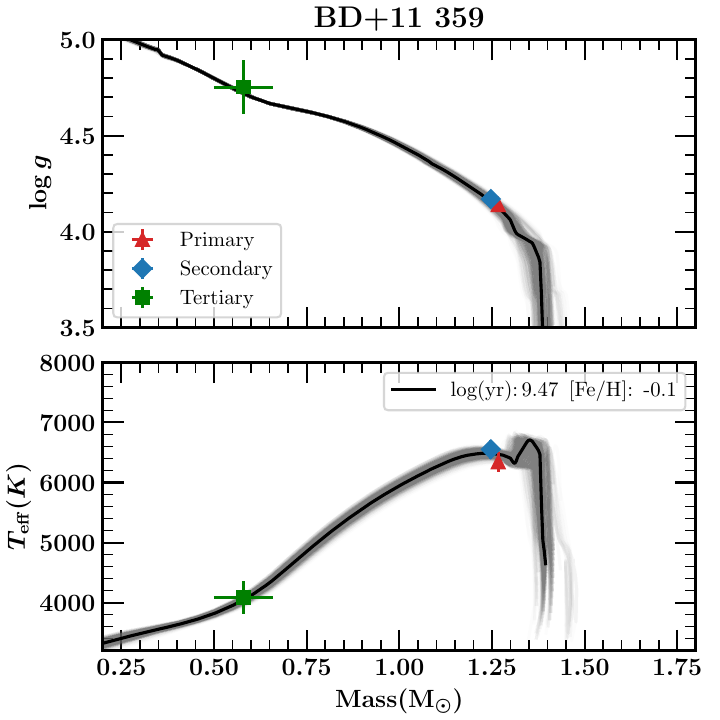}
    \caption{Same as Fig.~\ref{fig:CD58isofit} but for BD11.}
    \label{fig:BD11isofit}
\end{figure}

BD11 was the first CHT identified in the CR\'EME survey, already appearing (but not identified by name) in Figure~2 of \cite{litomysl_kris}. In the solutions by \citet{krisasas1} and \citet{bachestan}, one can note systematic effects in RV residuals. In those works, only a few RV measurements were used, and the system was treated as a `lone' binary; thus, our current results are more reliable. The inner binary is composed of two main-sequence stars of sub-solar metallicity and similar, yet unequal, masses, radii, and temperatures. We obtained a very good isochrone fit for the age of 2.95~Gyr (\autoref{fig:BD11isofit}). The fit predicts the tertiary to be a 0.58~M$_\odot$, 0.54~R$_\odot$, 4085~K dwarf. The mass is also close to the lower limit from RVs, which suggests an edge-on outer orbit and possibly a co-planar geometry.

\begin{table*}
 \centering
 \caption{Adopted values of all major parameters for the studied ST3 systems.  }
 \label{tab:paramtablest3}
 \small
\begin{tabular}{@{}lllllll}
\hline
 & \multicolumn{3}{c}{CD-32 6459} & \multicolumn{3}{c}{CD-62 1257} \\
\hline
\hline
\multicolumn{7}{c}{Orbital Parameters} \\
\hline
   & \multicolumn{2}{c}{Aa--Ab} & A--B  & \multicolumn{2}{c}{Aa--Ab} & A--B  \\
  \hline
  $t_0$ [BJD - 2450000]& \multicolumn{3}{c}{8547.61072(1)} & \multicolumn{3}{c}{8654.922092(1)} \\
  $P$ [days] & \multicolumn{2}{c}{ 4.02170747(4)} & 1372.1218$^a$  & \multicolumn{2}{c}{2.714577(2)} & 441.615(110)  \\
  \vspace{0.1cm}
  $a$ [R$_\odot$] & \multicolumn{2}{c}{14.85(2)} & 802(10) & \multicolumn{2}{c}{11.17(3)} & 392(14) \\
  \vspace{0.1cm}
  $e$ & \multicolumn{2}{c}{0.221(2)} &0.2688(50) & \multicolumn{2}{c}{0.007(3)} & 0.2997(144)  \\
  $i$ [deg] & \multicolumn{2}{c}{85.50(5)} & 51.2(5) & \multicolumn{2}{c}{89.4(1)} & 73(4)\\
  $\omega$ [deg]& \multicolumn{2}{c}{ 130.0(4)} & 39(2)  & \multicolumn{2}{c}{129.5(7)} & 218(2) \\
  $q$ & \multicolumn{2}{c}{0.9324(7)} &0.36(2) & \multicolumn{2}{c}{0.904(1)} & 0.64(8) \\
 \hline
\multicolumn{7}{c}{Stellar and atmospheric parameters} \\
\hline
   & Aa & Ab &  B & Aa & Ab &  B \\
 $M$ [M$_\odot$] & 1.4058(7) & 1.3107(9) & 0.97(5)&1.331(1) & 1.2025(9) & 1.6(2)\\
 \vspace{0.1cm}
 $R$ [R$_\odot$] &  1.57(1) & 1.44(1) & $0.9(2)^c$ & 1.59(2) & 1.43(3) & 1.9(2) \\
 
 $T_\mathrm{eff}$ [K]& 6674(248)  &6295(244) & 6344(412) &6729(263)  &6525(262)   & 6623(143) \\
 $\log(g)$ [dex] & 4.209(8) & 4.254(9) & 3.8(5)& 4.17(1) & 4.23(2) & 4.1(1)\\
 $v\mathrm{sin}(i)$ [km~s$^{-1}$] & 15(2) & 11(3) &4(4) & 30(2) &27(2) & 18(1) \\
 $\alpha$  [dex]    &0.12(6) & -0.28(1) &0.0$^a$ & 0.00(8)   &-0.05(9)& 0.00(5) \\
 \hline
\multicolumn{7}{c}{System parameters} \\
  \hline
  \vspace{0.1cm}
$\log(age)^c$ [dex] &\multicolumn{3}{c}{9.2(2)} & \multicolumn{3}{c}{9.39(3)} \\
  \vspace{0.1cm}
$\mathrm{[M/H]}_\mathrm{iSpec}$  [dex]   &\multicolumn{3}{c}{-0.25(3)} & \multicolumn{3}{c}{0.27(4)} \\
  \vspace{0.1cm}
$\mathrm{[Fe/H]}^c_\mathrm{isoc}$  [dex]   &\multicolumn{3}{c}{-0.1(2)} & \multicolumn{3}{c}{0.09(6)} \\
\vspace{0.1cm}
$E(B-V)^{c}$ [mag]    &\multicolumn{3}{c}{0.11(9)} & \multicolumn{3}{c}{0.03(2)} \\
Distance$^{c}$ [pc]   &\multicolumn{3}{c}{389(37)} & \multicolumn{3}{c}{432(11)} \\  
\hline
\end{tabular}

$^a$ Fixed during optimisation. $^b$ Obtained from empirical tables.  $^c$ Based on isochrone fitting. 

\end{table*}


\begin{table*}
 \centering
 \caption{Same as \autoref{tab:paramtablest3} but for ST2 systems.}
 \label{tab:paramtablest2}
 \small
\begin{tabular}{@{}lllllll}
\hline
 & \multicolumn{3}{c}{BD+11 359} & \multicolumn{3}{c}{CD-58 963} \\
\hline
\hline
\multicolumn{7}{c}{Orbital Parameters} \\
\hline
   & \multicolumn{2}{c}{Aa--Ab} & A--B  & \multicolumn{2}{c}{Aa--Ab} & A--B  \\
  \hline
  $T_0$ [BJD - 2450000]& \multicolumn{3}{c}{9448.20344(1)} & \multicolumn{3}{c}{9989.8842245(7) } \\
  $P$ [d] & \multicolumn{2}{c}{ 3.604795(2)} & 168.581(305)  & \multicolumn{2}{c}{3.5520444(2)} & 76.319(169)\\
  \vspace{0.1cm}
  $a$ [R$_\odot$] & \multicolumn{2}{c}{13.46(1)} & 184.16$^d$ & \multicolumn{2}{c}{12.78(3)} & 105.43$^d$ \\
  \vspace{0.1cm}
  $e$ & \multicolumn{2}{c}{0.0017(2)} & 0.0250(209) & \multicolumn{2}{c}{0.02(2)} & 0.2148(96) \\
  $i$ [deg] & \multicolumn{2}{c}{85.95(2)} & - & \multicolumn{2}{c}{81.58(6)} & - \\
  $\omega$ [deg]& \multicolumn{2}{c}{ 90.6(5)} & 359(36) & \multicolumn{2}{c}{179(7)} & 233(3)  \\
  $q$ & \multicolumn{2}{c}{0.984(4)} &0.173(3)& \multicolumn{2}{c}{0.92(8)} &0.22(1) \\
 \hline
\multicolumn{7}{c}{Stellar and atmospheric parameters} \\
\hline
   & Aa & Ab &  B & Aa & Ab &  B \\
 $M$ [M$_\odot$] & 1.267(4) &1.247(4) &0.58(8)$^c$&1.16(1)& 1.06(1) & 0.45(2)$^c$\\
 \vspace{0.1cm}
 $R$ [R$_\odot$] &  1.609(8) &1.554(8) &0.54(8)$^c$& 1.37(6)& 1.16(6) & 0.42(1)$^c$ \\
 
 $T_\mathrm{eff}$ [K]& 6342(181) & 6553(178) & 4085(272)$^c$ & 6478(171) & 6372(177)  & 3732(286)$^c$\\
 $\log(g)$ [dex] &4.143(5) &  4.167(5) & 4.7(1)$^c$ &4.25(4) & 4.35(6) & 4.86(5)$^c$ \\
 $v\mathrm{sin}(i)$ [km~s$^{-1}$] & 23.90(39) & 21.54(43) & - & 20(1) &18(1) & - \\
 $\alpha$  [dex]    &-0.13(3)& 0.08(2) & - & 0.05(6)  &0.08(4)  & - \\
 \hline
\multicolumn{7}{c}{System parameters} \\
  \hline
  \vspace{0.1cm}
$\log(age)^c$ [dex] &\multicolumn{3}{c}{9.47(1)} & \multicolumn{3}{c}{9.49(1)} \\
  \vspace{0.1cm}
$\mathrm{[M/H]}_\mathrm{iSpec}$  [dex]   &\multicolumn{3}{c}{-0.12(1)} & \multicolumn{3}{c}{-0.31(4)} \\
  \vspace{0.1cm}
$\mathrm{[Fe/H]}^c_\mathrm{isoc}$  [dex]   &\multicolumn{3}{c}{-0.10(4)} & \multicolumn{3}{c}{-0.15(6)} \\
\vspace{0.1cm}
$E(B-V)^{c}$ [mag]    &\multicolumn{3}{c}{0.09(1)} & \multicolumn{3}{c}{0.056(1)} \\
Distance$^{c}$ [pc]   &\multicolumn{3}{c}{253(2)} & \multicolumn{3}{c}{339(7)} \\  
\hline
\end{tabular}

$^a$ Fixed during optimisation. $^b$ Obtained from empirical tables.  $^c$ Based on isochrone fitting. $^d$ From \rebound{}

\end{table*}

\section{Distributions of parameters from detailed CHT solutions}

While spectroscopy is important in extracting precise and accurate parameters of stars in E2CHT, many such detailed solutions have been obtained using E3CHT and photodynamical modelling of long-term LCs. Such studies in the literature have estimated the metallicity of the systems (and sometimes of the individual stars) with the help of spectral energy distributions and isochrone fitting. While some part of the methodology assumes co-evolution and is model-dependent, it contributes substantially to the understanding of these rare systems. Going through the literature, we found 48 systems that have undergone such an analysis; these are listed in \autoref{tab:CHTlit}. To check for hints of formation and evolution channels, we looked at statistics of all the well-studied systems. We added our measurements from \cite{moharanacht,moharanasol} and this work. 
\subsection{Metallicity dependence of parameter distributions}
\begin{figure*}
    \centering
    \includegraphics[width=0.8\textwidth]{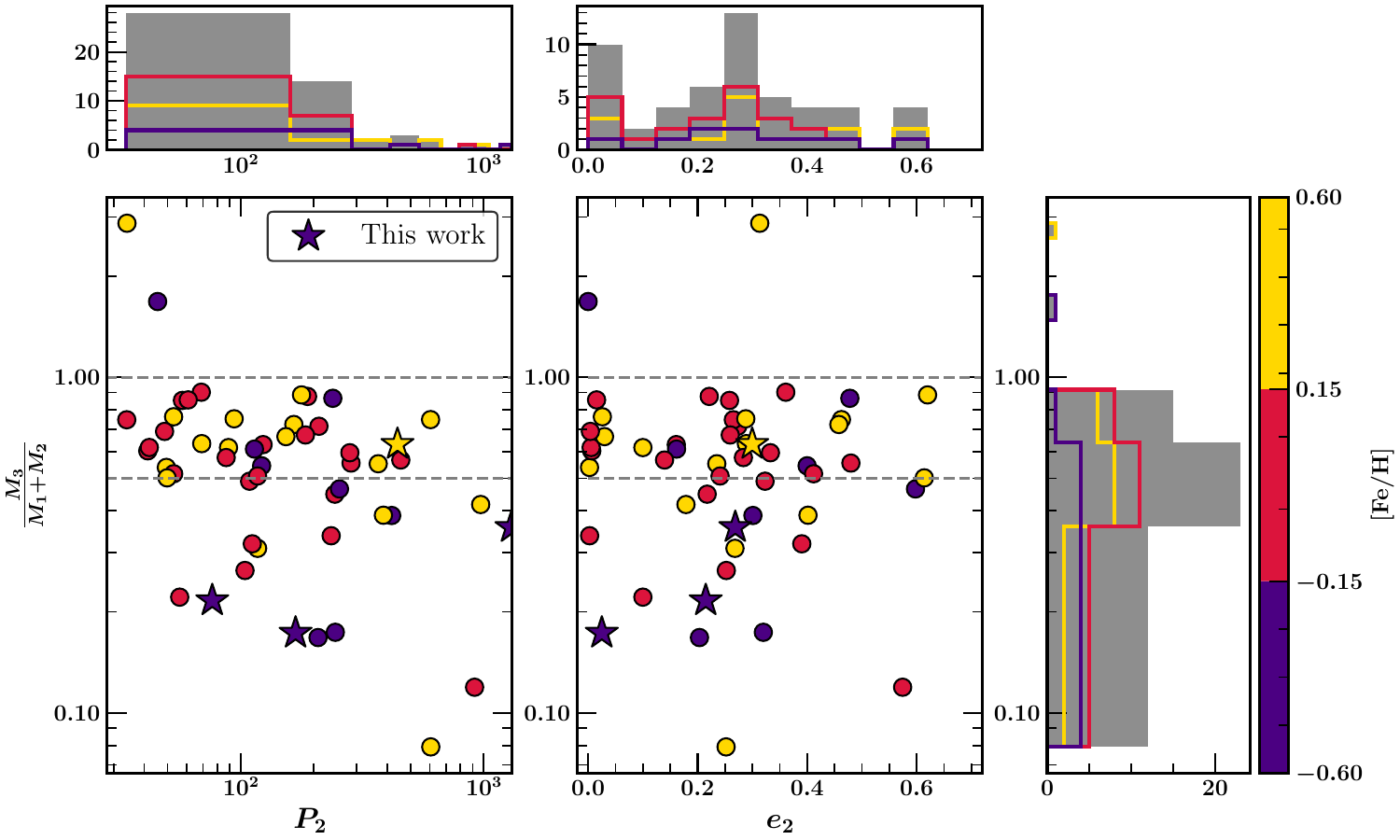}
    \caption{Distribution of the tertiary mass-ratio ($\frac{M_3}{M_1+M_2}$), the tertiary period ($P_2$), and the eccentricity ($e_2$) of the outer orbit for CHTs with metallicity estimates. The left panel shows the distribution of $\frac{M_3}{M_1+M_2}$ and $P_2$.  The right panel shows the distribution of $\frac{M_3}{M_1+M_2}$ with $e_2$ of the outer orbit. The points have colours according to their systemic metallicity, where metal-poor ($<$-0.15 dex) systems are purple, near solar-metallicity (-0.15 dex$<$[Fe/H]$<$0.15 dex) ones are red, and metal-rich ($>$0.15 dex) CHTs are yellow. The narrow panels on each axes represent the histograms of the respective parameters, with the coloured histograms representing the distributions for each metallicity group.}
    \label{fig:metdist}
\end{figure*}

\begin{figure}
    \centering
    \includegraphics[width=0.49\columnwidth]{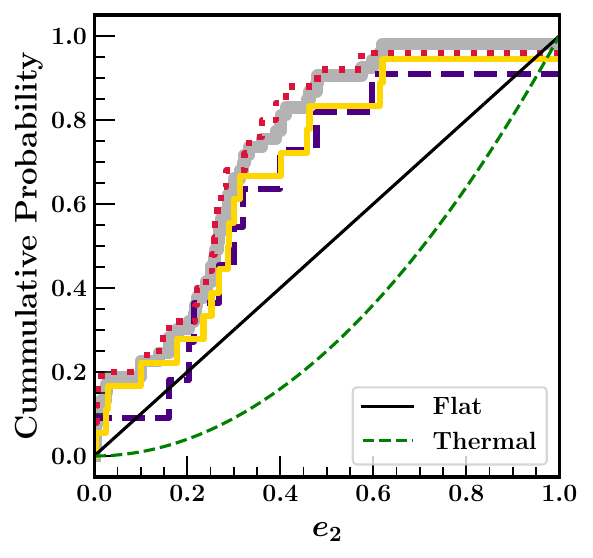}
    \includegraphics[width=0.49\columnwidth]{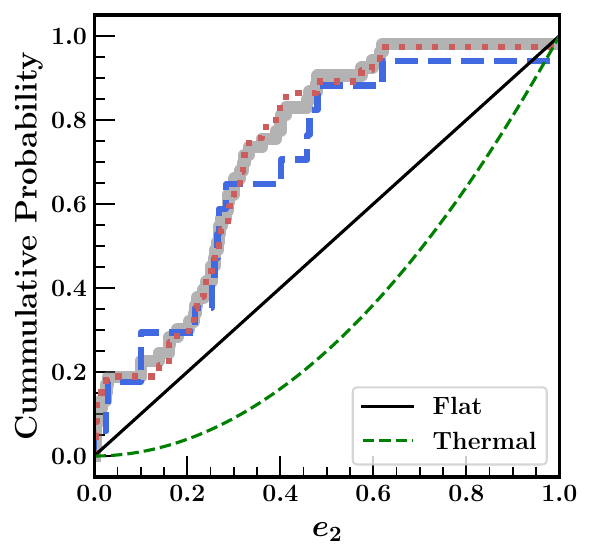}
    \caption{Cumulative eccentricity ($e_2$) distribution of the tertiary orbit of CHTs listed in \autoref{tab:CHTlit}. Left: Cumulative $e_2$ distribution of CHTs of different [fe/H] values. Yellow represents metal-rich samples, red represents solar ones, and purple represents metal-poor ones. Right: Cumulative $e_2$ distribution for young (blue) and old (red) systems. The black line represents the expected trend for a flat distribution and the dashed green line is the expected trend for a thermal distribution of $e_2$. The grey lines in both panels show the total distribution of all the listed CHTs.}
    \label{fig:eccdist}
\end{figure}

The distributions of the outer eccentricity, $e_2$, the tertiary period, $P_2$, and the tertiary to binary mass ratio ($\frac{M_3}{M_1+M_2}$) as a function of metallicity ([Fe/H]) are shown in \autoref{fig:metdist}. We divided the [Fe/H] into three regions (to account for any errors in measurements): sub-solar ($<$-0.15~dex), solar (-0.15 to +0.15~dex), and above-solar ($>$+0.15~dex).  The $P_2$ distribution peaks below 200 days, which is likely a bias due to the majority of the systems being E3CHT. The lack of low $\frac{M_3}{M_1+M_2}$ for periods of less than 200 days is suggestive of the brown-dwarf desert analogue \citep{bddesert2}, which we find to be [Fe/H]-independent. The distributions of $\frac{M_3}{M_1+M_2}$ for solar and above-solar metallicities both peak between 0.5 and 1, which are possible values for scenarios in which all three masses are equal, and the tertiary and binary masses are equal, respectively. However, the sub-solar systems seem to be spread uniformly. The two solar and above-solar systems at the low $\frac{M_3}{M_1+M_2}$ end are EDV systems and one of them has [Fe/H] very close to our limits for being metal-poor ([Fe/H]=-0.12). Interestingly, in a study of close binaries, \cite{bate2019} found that metal-poor environments have low opacities
and high cooling rates of dense gas, which enhances small-scale fragmentation in the star-forming cloud.  Objects of all [Fe/H] values show a favoured outer eccentricity of 0.3, and some E3CHT systems contribute to a peak at 0. This peak was observed in the \Kepler{} sample of CHTs in \cite{borkovitskeptrip}, \cite{ogletrip}, and \cite{Czava2023}. We computed the cumulative distribution of $e_2$ for all [Fe/H] values and found it to be similar to the above studies, which show a non-flat and non-thermal $e_2$ distribution (see left panel of \autoref{fig:eccdist}). This eccentricity distribution was also observed in field binaries, as was noted by \cite{multiplicity}. The lower [Fe/H]  distribution shows signs of a flat distribution around 0.2 but this could be the result of an incomplete, low-number statistic.

\begin{figure*}
    \centering
    \includegraphics[width=0.8\textwidth]{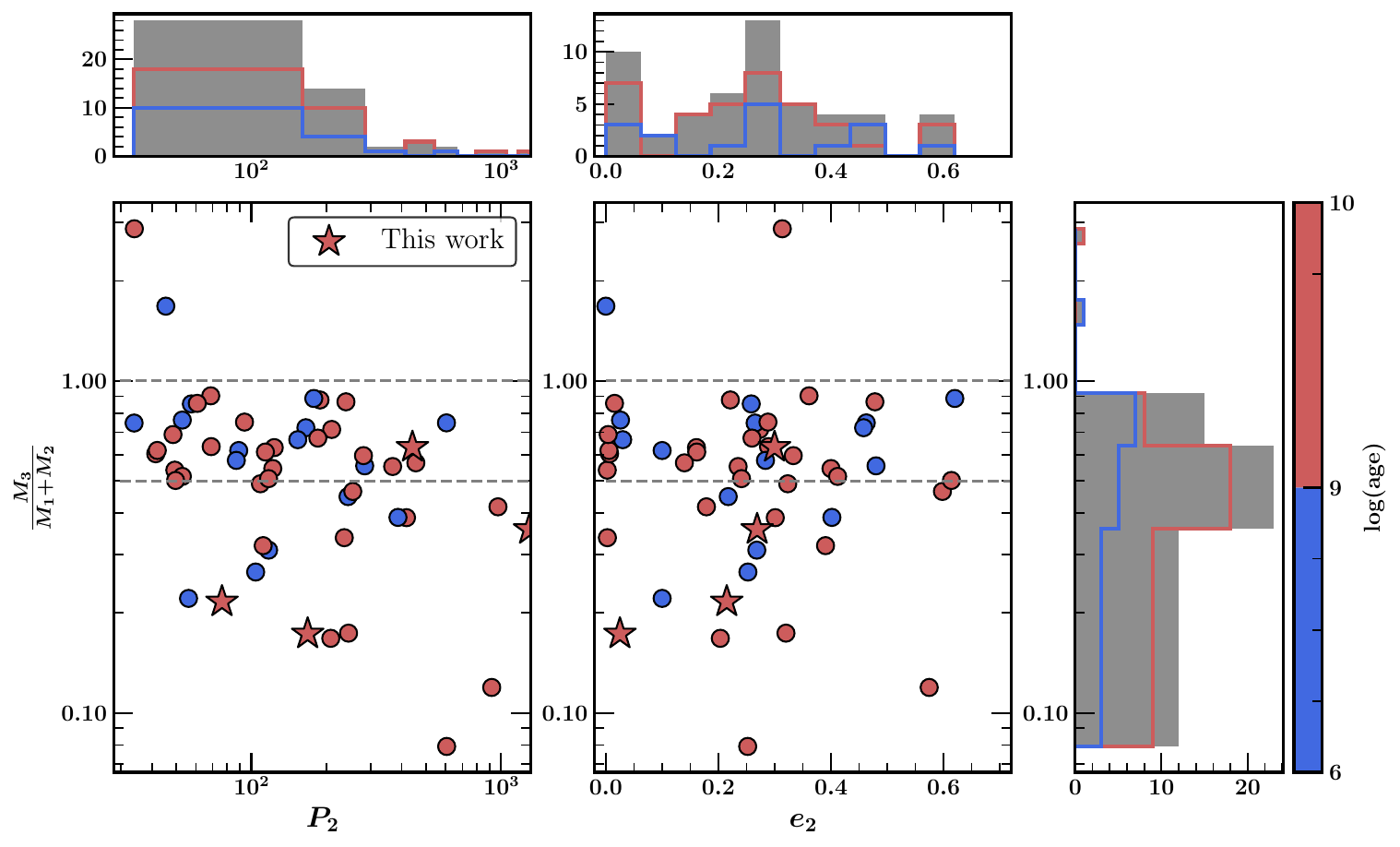}
    \caption{Same as \autoref{fig:metdist} but systems are marked with colours according to their age estimate (in log scale). The systems in red are old systems with $\log(age)>9$ (in Gyrs), while the systems in blue are classified as old with $\log(age)<9$.}
    \label{fig:agedist}
\end{figure*}

\subsection{Age dependence of parameter distributions}
\begin{figure}
    \centering
    \includegraphics[width=0.95\columnwidth]{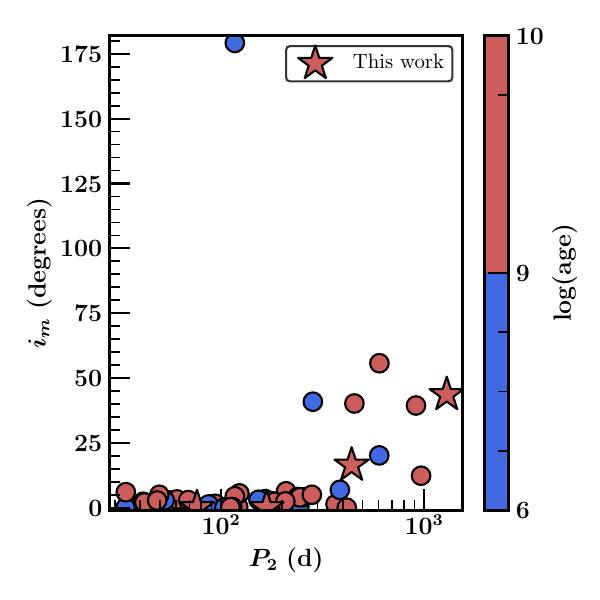}
    \caption{Distribution of mutual inclination (\incmut) vs period of the outer orbit ($P_2$). The blue circles represent young systems, while the red circles represent old systems. Stars denote systems observed in this work.}
    \label{fig:p2vsim}
\end{figure}

To see any signatures of dynamical changes in these distributions over time, we differentiated them based on the system age.  Due to the lack of very young systems (log(age)$\sim$6), we made a separation between young and old systems at a log(age) of 9. We see similar $e_2$ distributions with no drastic differences between young and old systems (see right panel of \autoref{fig:eccdist}). Meanwhile, the young systems favoured a mass ratio closer to 1 than the old systems, which peaked at around 0.5 (\autoref{fig:agedist}).  The change in the peak mass ratio between young and old systems can be a result of the migration of the tertiary. To see any signatures of migration of the tertiary, we plot the age dependency of the \incmut\footnote{For systems with an estimated range of \incmut{}, we consider the lower limit only} in \autoref{fig:p2vsim}. The distribution shows all old systems, below a period of 500 d, to be planar. All young systems below a period of 300 d are planar. While there is a young system with a high value of \incmut{}, it is still planar, though in a retrograde orbit. These \incmut-$P_2$ cut-offs are similar to the eccentricity-period cut-offs seen in observations of binaries in the solar neighbourhood \citep{solarmultiplesurv}. Simulations of close binary formation attribute this to tidal dissipation \citep{maxwellkratter}. The simulations also predict that the cut-offs move towards longer periods as the systems get old, similar to the trends seen in \autoref{fig:p2vsim}. This planarisation of CHTs may be powered by tidal dissipation (possibly KCTF), along with mass loss, which can explain the different mass ratios for old and young systems. The other plausible scenario is that lower-mass-ratio systems are dynamically more stable and hence survive longer.

Considering the current state of trends, it is possible that the outer hierarchical orbit of a CHT is formed in a similar way to that of close, field binaries. While the dependencies of these distributions are prone to error because of the small sample, it is an interesting property to study that could help us better understand star formation at these scales.

\section{Conclusions}

In this work, we present an analysis of three newly discovered CHTs and one already identified. We used LC modelling, RV modelling, spectral disentangling, and spectral analysis to get the orbital, stellar, and atmospheric parameters of all three stars in ST3 systems and two stars in the inner binary of our ST2 systems. For ST3 systems, we used the parameters of all three stars to constrain their ages and distances. For ST2 systems, we used parameters of the inner binary to estimate the ages and then estimate the tertiary mass and radius with constraints from the estimated minimum mass and {\it Gaia} distances. All inner binaries have their masses estimated with high or very high precision. Except for CD58 Aa and Ab, the radii are precise to a level of 1\% or better. The four systems are summarised below:

\begin{itemize}
    \item CD-32 6459: This system is the widest in our sample. The tertiary star is a 0.97 $M_{\odot}$ main-sequence star. This is a metal-poor system with an age of 1.58 Gyr. The inner binary is eccentric and has a similar eccentricity to the tertiary orbit ($\sim$0.2). The mutual inclination limits are close to the limiting angle for ZLK oscillations, which could explain the large inner eccentricity. This system is a candidate to observe KCTF in process. 
    
    \item CD-62 1257: This is a system with a larger and more massive tertiary than the stars in the eclipsing pair. The tertiary will evolve below the Roche limit of the A-B orbit, and the system will undergo a TCE phase. 

    \item CD-58 963: This system has the shortest outer period and lowest mass in our sample. The inner binary, composed of 1.16 and 1.06~M$_\odot$ stars, is accompanied by a 0.45~M$_\odot$ red dwarf. Despite it previously being suggested that this is a young ($\sim$20~Myr) system \citep{Borko2020}, we found a good isochrone fit for an older age ($\sim$3~Gyr).
    
    \item BD+11 359: This system is the first CHT identified by the CR\'EME survey. The inner binary, composed of 1.27 and 1.25~M$_\odot$ stars, is accompanied by a 0.58~M$_\odot$ red dwarf.
\end{itemize}

We compiled CHT systems from the literature with age and metallicity estimates and used them with our results to look at the dependencies of the tertiary mass ratio and eccentricity. We found that metal-poor stars have no preferred mass ratio but metal-rich and solar-metallicity stars prefer mass ratios around 0.5. Such a dependency of mass ratios upon metallicity was observed in simulations of close binaries. We also found that older systems ($>$1 Gyr) also prefer a mass ratio of around 0.5. Older CHTs with outer periods less than 500d and younger systems with outer periods less than 300d are near planar in configuration. The outer eccentricities of all CHTs follow the same distribution that was observed previously for CHTs, and also field binaries. This is suggestive of a CHT formation scenario similar to that of close, field binaries. However, these distributions are biased due to the abundance of triply eclipsing systems. They also suffer from a lack of systems, and this is why there is a need for more homogeneous and detailed studies of CHT parameters in the future.

\begin{acknowledgements}

The authors thank Dr. Tomer Shenar for his help with using his disentangling code, \textsc{dsaa}.
This work is funded by the Polish National Science Centre (NCN) through grant 2021/41/N/ST9/02746. K.G.H. acknowledges support from NCN grant 2023/49/B/ST9/01671.
A.M., F.M., T.P., and M.K. are supported by NCN through grant 2017/27/B/ST9/02727. F.M. gratefully acknowledges support from the NASA TESS Guest Investigator grant 80NSSC24K0498 (PI F.~Marcadon). 

Observations for CD-62 1257, CD-32 6459, and CD-58 963 were obtained with the Southern African Large Telescope (SALT). Polish participation in SALT is funded by grant No. MEiN nr 2021/WK/01. 
This work is based on observations collected at the European Southern Observatory, Chile under programmes 088.D-0080, 089.D-0097, 089.C-0415, 090.D-0061, 091.D-0145, and through CNTAC proposals 2012B-036, 2013A-093, 2013B-022, 2014B-067, and 2015A-074.

This paper includes data collected with the TESS mission, obtained from the Mikulski Archive for Space Telescopes (MAST) data archive at the Space Telescope Science Institute (STScI). Funding for the TESS mission is provided by the NASA Explorer Program. STScI is operated by the Association of Universities for Research in Astronomy, Inc., under NASA contract NAS 5–26555. This work also presents results from the European Space Agency (ESA) space mission {\it Gaia}. {\it Gaia} data are being processed by the {\it Gaia} Data Processing and Analysis Consortium (DPAC). Funding for the DPAC is provided by national institutions, in particular, the institutions participating in the {\it Gaia} MultiLateral Agreement (MLA). 
\end{acknowledgements} 
%
%
\bibliographystyle{aa} 
\bibliography{ECHT.bib} 
\begin{appendix} 
\section{RV measurements}
\label{appx:rvs}

We present all RV measurements used in this work in Tables~\ref{tab:rvst3} and \ref{tab:rvst2}. We included the data from \cite{krisasas1} for BD11. When a measurement cannot be done at a certain epoch due to various effects, for example line blending, it is marked with `---'. There is no data, obviously,  for the tertiaries in ST2 systems.

\begin{table*}
    \centering
    \caption{RV measurements of ST3s used in this work. All values are given in km s$^{-1}$.}\label{tab:rvst3}
    \begin{tabular}{c|rr|rr|rr|rr|c}
    \hline \hline
    JD-2400000 & $v_{Aa}$ & $\pm$ & $v_{Ab}$ & $\pm$ & $v_{COM}$ & $\pm$ & $v_{B}$ & $\pm$ & Inst. \\
    \hline
    \multicolumn{10}{c}{CD-32 6459}\\
    \hline
56380.736718 & -104.218 & 1.572 &  101.563 & 0.370 &  -4.919 & 0.871 &  16.050 & 0.371 & FEROS\\ 
56382.647745 &   68.482 & 0.615 &  -80.149 & 0.312 &  -3.240 & 0.463 &  16.352 & 0.532 & FEROS\\
56383.509933 &   50.556 & 1.284 &  -65.791 & 0.221 &  -5.587 & 0.654 &  17.108 & 0.817 & FEROS\\
56429.529095 &  -49.181 & 0.779 &   42.118 & 0.331 &  -5.125 & 0.546 & ---     & ---   & FEROS\\
57078.571717 &   81.145 & 0.915 &  -75.414 & 0.409 &   5.598 & 0.653 & -12.875 & 0.408 & CORALIE\\
59282.657460 &   72.450 & 1.120 &  -91.378 & 0.279 &  -6.605 & 0.637 &  22.575 & 0.221 & CHIRON\\
59294.641417 &   73.227 & 1.531 &  -90.608 & 0.299 &  -5.831 & 0.806 &  22.550 & 0.163 & CHIRON\\
59533.834998 &  -88.034 & 1.066 &  101.477 & 0.344 &   3.414 & 0.660 &  -2.010 & 0.180 & CHIRON\\
59556.828885 &   32.934 & 1.394 &  -29.043 & 0.351 &   3.027 & 0.791 &  -4.056 & 0.176 & CHIRON\\
59668.562841 &   82.137 & 0.628 &  -76.972 & 0.351 &   5.359 & 0.504 & -10.454 & 0.243 & CHIRON\\
59702.500783 &  -99.376 & 0.997 &  118.914 & 0.325 &   5.959 & 0.621 & -11.312 & 0.205 & CHIRON\\
59984.703358 &  -40.991 & 0.929 &   55.012 & 0.357 &   5.335 & 0.621 & -10.066 & 0.159 & CHIRON\\
59986.671959 &   82.311 & 0.492 &  -75.945 & 0.338 &   5.945 & 0.444 &  -9.947 & 0.235 & CHIRON\\
59989.631343 &   47.508 & 0.765 &  -40.108 & 0.331 &   5.229 & 0.546 &  -9.894 & 0.248 & CHIRON\\
60005.651318 &   40.960 & 0.601 &  -34.547 & 0.227 &   4.524 & 0.408 &  -9.627 & 0.311 & CHIRON\\
60020.659589 &  -67.240 & 0.724 &   83.724 & 0.312 &   5.607 & 0.519 &  -8.826 & 0.166 & CHIRON\\
60036.575983 &  -84.606 & 0.765 &  101.972 & 0.383 &   5.427 & 0.572 &  -8.180 & 0.328 & CHIRON\\
60037.549856 &   19.233 & 0.820 &   -9.177 & 0.266 &   5.524 & 0.513 & ---     & ---   & CHIRON\\
60067.588422 &   35.446 & 0.833 &  -28.408 & 0.318 &   4.633 & 0.557 &  -8.075 & 0.266 & CHIRON\\
60097.535108 &  -15.504 & 1.038 &   26.900 & 0.546 &   4.958 & 0.807 & ---     & ---   & CHIRON\\
59955.409682 &  -52.586 & 1.503 &   66.473 & 0.513 &   4.866 & 0.954 & -11.266 & 0.363 & HRS\\
59979.590299 &  -62.979 & 1.544 &   77.115 & 0.708 &   4.623 & 1.116 & -10.709 & 0.229 & HRS\\
60012.493571 &  -80.761 & 1.476 &   96.673 & 0.812 &   4.859 & 1.163 &  -9.380 & 0.419 & HRS\\
60028.470978 &  -91.177 & 1.257 &  108.103 & 0.617 &   4.985 & 0.935 &  -8.409 & 0.468 & HRS\\
\hline
    \multicolumn{10}{c}{CD-62 1257}\\
    \hline
56100.669762 &  -98.916 & 1.822 &   39.278 & 2.324 & -33.287 & 2.060 & -11.280 & 0.620 & FEROS \\
56100.806359 & -118.969 & 2.095 &   67.377 & 2.282 & -30.472 & 2.184 & -11.980 & 0.548 & FEROS \\
56102.604977 &   55.178 & 1.801 & -129.758 & 1.190 & -32.649 & 1.511 & -12.250 & 0.291 & FEROS \\
56102.658367 &   45.432 & 2.042 & -116.760 & 0.993 & -31.594 & 1.544 & -11.946 & 0.493 & FEROS \\
56193.656641 &  -87.689 & 1.812 &   62.349 & 1.433 & -16.435 & 1.632 & -32.131 & 0.462 & FEROS \\
56194.660546 &   79.540 & 1.888 & -126.252 & 1.601 & -18.192 & 1.752 & -32.916 & 0.600 & FEROS \\
56517.612517 &   57.381 & 2.069 & -136.802 & 2.325 & -34.838 & 2.191 &  -7.521 & 1.257 & FEROS \\
56520.767555 &   39.464 & 2.434 & -113.775 & 1.816 & -33.310 & 2.141 &  -7.982 & 0.378 & FEROS \\
56729.869412 &   63.613 & 0.869 &  ---     & ---   &   ---   & ---   & ---     & ---   & CORALIE \\
56730.887230 &  ---     &   --- &  111.925 & 2.518 &   ---   & ---   & ---     & ---   & CORALIE \\
56939.497908 &  -98.923 & 1.023 &  ---     & ---   &   ---   & ---   & ---     & ---   & CORALIE \\
56941.505155 &   42.437 & 1.797 & -122.590 & 2.568 & -35.935 & 2.163 &  -5.126 & 0.626 & CORALIE \\
57181.628870 &  ---     &   --- &   67.125 & 2.640 &   ---   & ---   & ---     & ---   & CORALIE \\
57182.619647 &   85.974 & 3.040 &  ---     & ---   &   ---   & ---   & ---     & ---   & CORALIE \\
59336.827822 &  -98.399 & 1.529 &   96.284 & 3.654 &  -5.943 & 2.538 & -50.386 & 0.289 & CHIRON \\
59340.897524 &   83.719 & 1.435 & -106.865 & 2.830 &  -6.791 & 2.097 & -51.592 & 0.366 & CHIRON \\
59344.821991 &  ---     &   --- &   78.432 & 2.119 &   ---   & ---   & ---     & ---   & CHIRON \\
59344.920517 &  ---     &   --- &   91.702 & 1.100 &   ---   & ---   & ---     & ---   & CHIRON \\
59345.907492 &   16.419 & 1.327 &  ---     & ---   &   ---   & ---   & ---     & ---   & CHIRON \\
59673.888987 & -109.688 & 0.918 &   57.383 & 1.820 & -30.345 & 1.346 & -18.628 & 0.217 & CHIRON \\
59674.874533 &   67.968 & 1.030 & -134.531 & 2.150 & -28.200 & 1.562 & -18.659 & 0.259 & CHIRON \\
59675.876626 &  -74.877 & 1.736 &  ---     & ---   &   ---   & ---   & ---     & ---   & CHIRON \\
59677.870202 &   63.546 & 1.144 & -128.554 & 0.803 & -27.684 & 0.982 & -19.592 & 0.196 & CHIRON \\
59737.680701 &   60.570 & 1.792 & -109.053 & 2.322 & -19.985 & 2.044 & -35.250 & 0.335 & CHIRON \\
60025.916463 &  -51.549 & 0.828 &  ---     & ---   &   ---   & ---   & ---     & ---   & CHIRON \\
60040.901851 &  ---     &   --- &  -58.550 & 2.718 &   ---   & ---   & ---     & ---   & CHIRON \\
59708.518169 &  -80.575 & 2.285 &   36.479 & 1.747 & -24.985 & 2.030 & -27.066 & 0.455 & HRS \\
59723.493185 &   45.771 & 1.374 &  -96.058 & 1.197 & -21.585 & 1.290 & -31.520 & 0.198 & HRS \\
59807.431589 &   18.130 & 1.218 &  ---     & ---   &   ---   & ---   & ---     & ---   & HRS \\
60021.630343 &  -69.462 & 1.362 &  ---     & ---   &   ---   & ---   &  -5.011 & 0.204 & HRS \\
60031.625282 & -106.984 & 1.228 &   37.348 & 1.740 & -38.440 & 1.471 &  -6.925 & 0.267 & HRS \\
60148.362576 &  -92.928 & 1.541 &   55.975 & 1.820 & -22.213 & 1.673 & -26.912 & 0.198 & HRS \\
60158.444732 &   60.291 & 1.448 & -119.168 & 1.421 & -24.935 & 1.435 & -29.744 & 0.138 & HRS \\
60169.312112 &   61.995 & 1.270 & -110.703 & 2.192 & -20.020 & 1.708 & -33.061 & 0.207 & HRS \\
    \hline
    \end{tabular}
\end{table*}

\begin{table*}
    \centering
    \caption{RV measurements of ST2s used in this work. All values are given in km s$^{-1}$.}\label{tab:rvst2}
    \begin{tabular}{c|rr|rr|rr|c}
    \hline \hline
    JD-2400000 & $v_{Aa}$ & $\pm$ & $v_{Ab}$ & $\pm$ & $v_{COM}$ & $\pm$ & Inst. \\
    \hline
    \multicolumn{8}{c}{CD-58 963}\\
    \hline
59803.615462 &   29.174 & 0.922 &  -35.492 & 0.707 &  -1.766 & 0.293 & HRS \\
59851.524928 &  -27.411 & 0.483 &   24.944 & 0.382 &  -2.361 & 0.156 & HRS \\
59861.456438 &   76.812 & 0.458 &  -73.061 & 0.572 &   5.103 & 0.184 & HRS \\
59872.465435 &   32.610 & 0.452 &  -30.733 & 0.503 &   2.303 & 0.171 & HRS \\
59893.370953 &   61.685 & 0.461 &  -95.980 & 0.631 & -13.752 & 0.194 & HRS \\
59894.400083 &  -73.674 & 0.461 &   48.467 & 0.648 & -15.234 & 0.197 & HRS \\
59913.360440 &  -49.362 & 0.553 &   -1.266 & 0.614 & -26.350 & 0.209 & HRS \\
59914.342234 &   59.785 & 0.567 & -118.984 & 0.726 & -25.750 & 0.230 & HRS \\
59920.309846 &  -60.173 & 0.564 &   34.240 & 0.665 & -15.000 & 0.219 & HRS \\
59946.440661 &   89.040 & 0.536 &  -89.817 & 0.648 &   3.463 & 0.211 & HRS \\
59984.377028 &  -53.007 & 0.444 &    1.721 & 0.642 & -26.822 & 0.193 & HRS \\
60031.260702 &   68.260 & 0.475 &  -75.713 & 0.732 &  -0.626 & 0.214 & HRS \\
60186.569821 &  -64.561 & 0.681 &   64.689 & 0.995 &  -2.720 & 0.298 & HRS \\
\hline
    \multicolumn{8}{c}{BD+11 359}\\
    \hline
55876.635630 &  -84.750 & 0.243 &  103.689 & 0.389 &   8.729 & 0.245 & FEROS \\
55878.625064 &   95.977 & 0.368 &  -81.745 & 0.372 &   7.814 & 0.287 & FEROS \\
55878.706579 &   91.328 & 0.602 &  -75.655 & 0.317 &   8.492 & 0.357 & FEROS \\
56193.796462 &  -80.139 & 0.472 &  107.504 & 0.521 &  12.945 & 0.384 & FEROS \\
56195.766059 &  105.649 & 0.405 &  -80.877 & 0.421 &  13.118 & 0.320 & FEROS \\
56290.562677 &  -42.229 & 0.432 &   41.854 & 0.565 &  -0.517 & 0.386 & FEROS \\
56291.565524 &  -74.618 & 0.281 &   76.734 & 0.497 &   0.463 & 0.301 & FEROS \\
56292.551540 &   66.484 & 0.349 &  -66.822 & 0.657 &   0.354 & 0.389 & FEROS \\
56517.837030 &  -51.987 & 0.825 &   79.353 & 0.833 &  13.166 & 0.643 & FEROS \\
56518.805909 &  -47.078 & 1.219 &   74.302 & 0.843 &  13.135 & 0.800 & FEROS \\
56519.828021 &   95.671 & 0.352 &  -70.041 & 0.558 &  13.466 & 0.352 & FEROS \\
56137.878521 &   95.782 & 0.308 &  -89.616 & 0.363 &   3.811 & 0.260 & HARPS \\
56179.805929 &  -61.640 & 0.362 &   90.403 & 0.332 &  13.784 & 0.269 & HARPS \\
56237.643240 &  -55.525 & 0.888 &   59.706 & 0.403 &   1.637 & 0.502 & CORALIE \\
56238.685091 &   86.698 & 0.746 &  -84.060 & 0.728 &   1.990 & 0.572 & CORALIE \\
56242.736340 &   87.777 & 0.952 &  -88.218 & 0.794 &   0.471 & 0.677 & CORALIE \\
56497.865091 &   56.496 & 0.715 &  -34.776 & 0.651 &  11.218 & 0.530 & CORALIE \\
56619.641008 &  -73.217 & 0.807 &   71.866 & 1.177 &  -1.246 & 0.768 & CORALIE \\
54727.175538 &  -58.281 & 0.781 &   59.466 & 0.624 &   0.129 & 0.545 & UCLES \\
54748.173087 &  -93.788 & 0.438 &   90.371 & 0.464 &  -2.432 & 0.349 & UCLES \\
54837.008123 &   85.826 & 0.659 &  -58.741 & 1.404 &  14.110 & 0.797 & UCLES \\
54838.029710 &  -58.900 & 1.149 &   88.605 & 0.679 &  14.272 & 0.710 & UCLES \\
54839.966130 &   97.981 & 0.699 &  -72.071 & 0.875 &  13.623 & 0.610 & UCLES \\
    \hline
    \end{tabular}
\end{table*}

\section{JKTEBOP tables}
\label{appx:jktebop}

We made LC models with \JKTEBOP{} for each TESS{} sector separately. The parameters obtained from the fitting for each fun are given in \autoref{tab:LC_CD32} (CD32), \autoref{tab:LC_BD11} (BD11), \autoref{tab:LC_CD62} (CD62), and \autoref{tab:LC_CD58} (CD58).  For our final estimates, we use only the solutions which have $J$ close to the spectroscopic value. Some sectors have nonphysical solutions due to strong effects of stellar activity. This also affects the estimation of other orbital parameters, \ie{}, high eccentricity in S62 due to high $J$ value (see \autoref{tab:LC_CD58}).

\begin{table}[]
    \centering
    \caption{JKTEBOP Solutions for CD32.}
    \label{tab:LC_CD32}
    \begin{tabular}{c|c|c}
    \hline
\hline
        Parameters  & S09  & S62 \\
        \hline
        $P$ [d]  & 4.021750(7) & 4.021758(4)\\
         $T_{0}$ [JD-2457000] &1547.610718(27)  & 2991.403699(14)\\
         $i$ [deg]  & 85.47(3)  &85.52(2) \\
         $e$  & 0.2254(11)  & 0.2174(8)\\
         $\omega$ [deg]  &129.1(2) & 130.9(2)\\
         $J$     &0.857(13) &0.791(9) \\
         $r_1+r_2$     &0.2033(2) &0.2034(2)\\
         $k$ &   0.932(4)&0.900(6) \\
         $L_{3}$ & 0.0936(33)  &0.1080(25) \\
         \hline
    \end{tabular}
        \vspace{0.2cm}
\end{table}

\begin{table}[]
    \centering
    \caption{JKTEBOP Solutions for BD11.}
    \label{tab:LC_BD11}
    \begin{tabular}{c|c|c}
        \hline
\hline
        Parameters & S42 & S43  \\
        \hline
        $P$ [d] &3.604797(3) &  3.604807(3) \\
         $T_{0}$ [JD-2457000]& 2448.203442(14) & 2477.041918(15) \\
          $i$ [deg]  &85.96(1) &  85.94(1) \\
          $e$  & 0.0013(1)&0.0021(2) \\
          $\omega$ [deg] &89.2(3) & 91.9(2)\\
          $J$  &  1.0034(2)  &1.0047(1) \\
          $r_1+r_2$  & 0.234940(3)   &0.23510(4) \\
          $k$ & 0.965(5) &  0.967(5) \\
          $L_{3}$ &0.0603(14) &0.0554(14)   \\
         \hline
         
    \end{tabular}
        \vspace{0.2cm}
\end{table}

\begin{table*}
    \centering
    \caption{JKTEBOP Solutions for CD62.}
        \label{tab:LC_CD62}
    \begin{tabular}{c|c|c|c}
        \hline
\hline
        Parameters & S13 & S66 & S67\\
        \hline
        $P$ [d] & 2.714600(4)& 2.714652(3) &2.714696(5)\\
         $T_{0}$ [JD-2457000]&1654.922099(2) & 3099.154856(20)&3129.016155(25)\\
          $i$ [deg]  & 89.53(6)&89.31(4)  &89.34(5)\\
          $e$  &0.0133(7) & 0.0031(7) &0.0044(8)\\
          $\omega$ [deg] &88.3(1) &83.2(16) &85.6(9)\\
          $J$  &  0.999(7)  &0.924(6) &0.943(8)\\
          $r_1+r_2$  &  0.2512(1)  &0.2528(1) &0.2527(1)\\
          $k$ & 0.825(1) & 0.814(1) &0.817(2)\\
          $L_{3}$ &0.5711(7) & 0.5730(7) &.5728(9)\\
         \hline
    \end{tabular} \\ 
\end{table*}

\begin{table*}
    \centering
    \caption{JKTEBOP Solutions for CD58.}
    \label{tab:LC_CD58}
    \begin{tabular}{c|c|c|c|c|c|c}
        \hline
\hline
Parameters & S62 & S63 & S64 & S67 & S68 & S69 \\
\hline
$P$ [d] & 3.55385(4) & 3.55137(4) & 3.55174(4) & 3.55173(8) & 3.55174(4) & 3.55155(4) \\
 $T_{0}$ [JD-2457000] & 2989.88257(8) & 3021.85611(8) & 3046.71543(8) & 3128.41124(9) & 3160.38653(8) & 3192.34897(8) \\
$i$ [deg]  & 81.93(1) & 81.63(2) & 81.59(1) & 81.58(2) & 81.56(1) & 81.60(2) \\
$e$ & 0.228(3) & 0.016(4) & 0.018(5) & 0.024(6) & 0.015(4) & 0.011(5) \\
$\omega$ [deg] & 90.25(1) & 94(1) & 266(1) & 92.3(6) & 95(2) & 265(3) \\
$J$ & 5.4(1) & 1.04(4) & 0.78(3) & 1.12(6) & 1.03(4) & 0.83(3) \\
$r_1+r_2$ & 0.2010(2) & 0.1969(3) & 0.1976(3) & 0.1977(3) & 0.1982(3) & 0.1974(3) \\
$k$ & 0.511(7) & 0.77(2) & 0.92(2) & 0.79(2) & 0.80(1) & 0.89(2) \\
$L_{3}$  &  0(fixed) &  0(fixed) &  0(fixed) &  0(fixed) &  0(fixed) &  0(fixed)   \\
         \hline
    \end{tabular}
\end{table*}

\section{\textsc{isofitter} results}
\label{appx:isofitter}
\begin{figure*}[h]
    \includegraphics[width=0.46\textwidth]{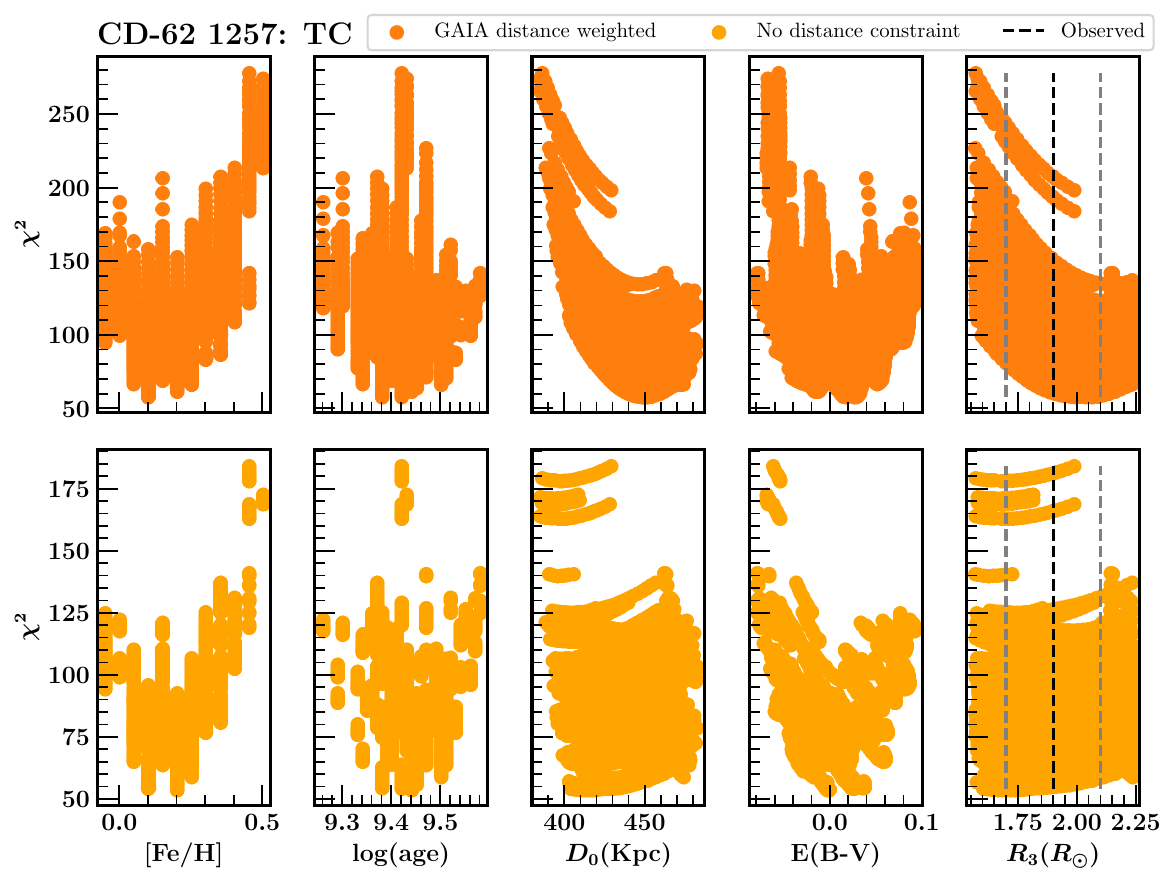}
    \includegraphics[width=0.46\textwidth]{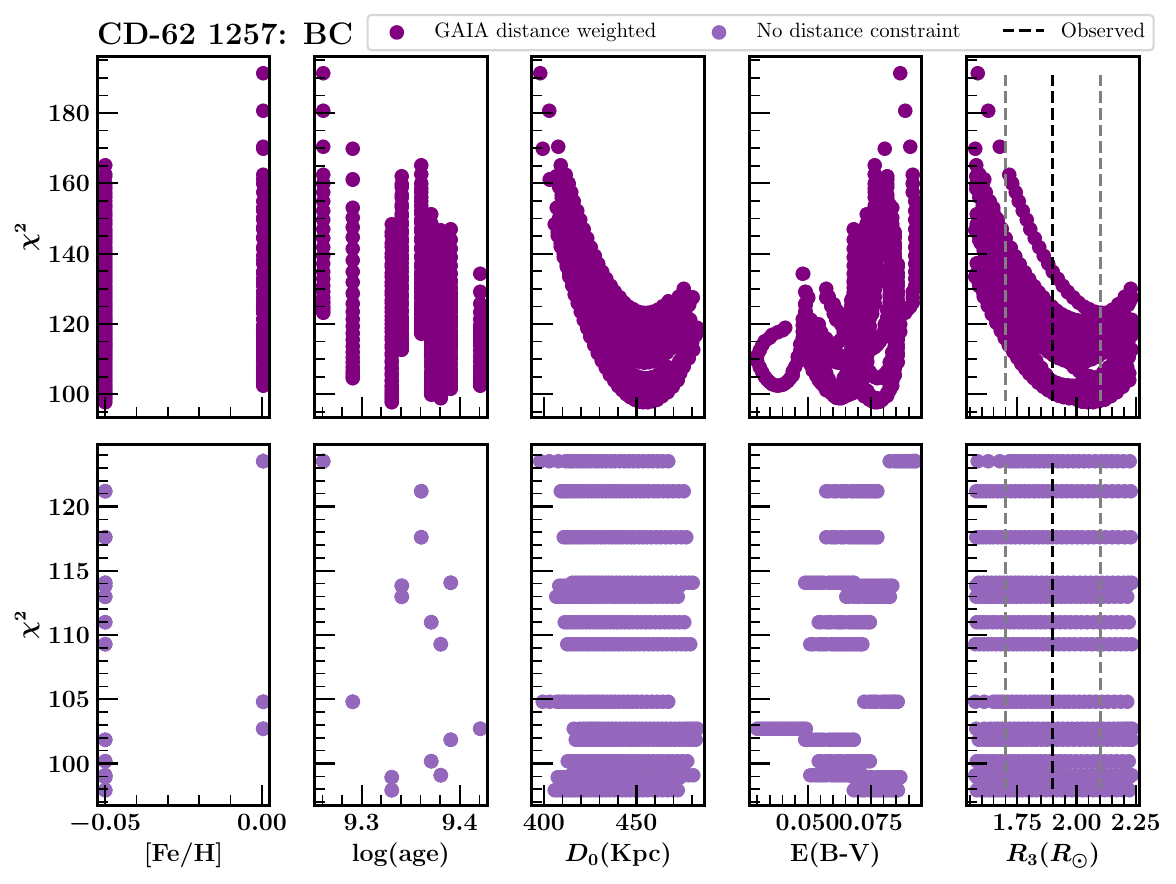} 
      \includegraphics[width=0.46\textwidth]{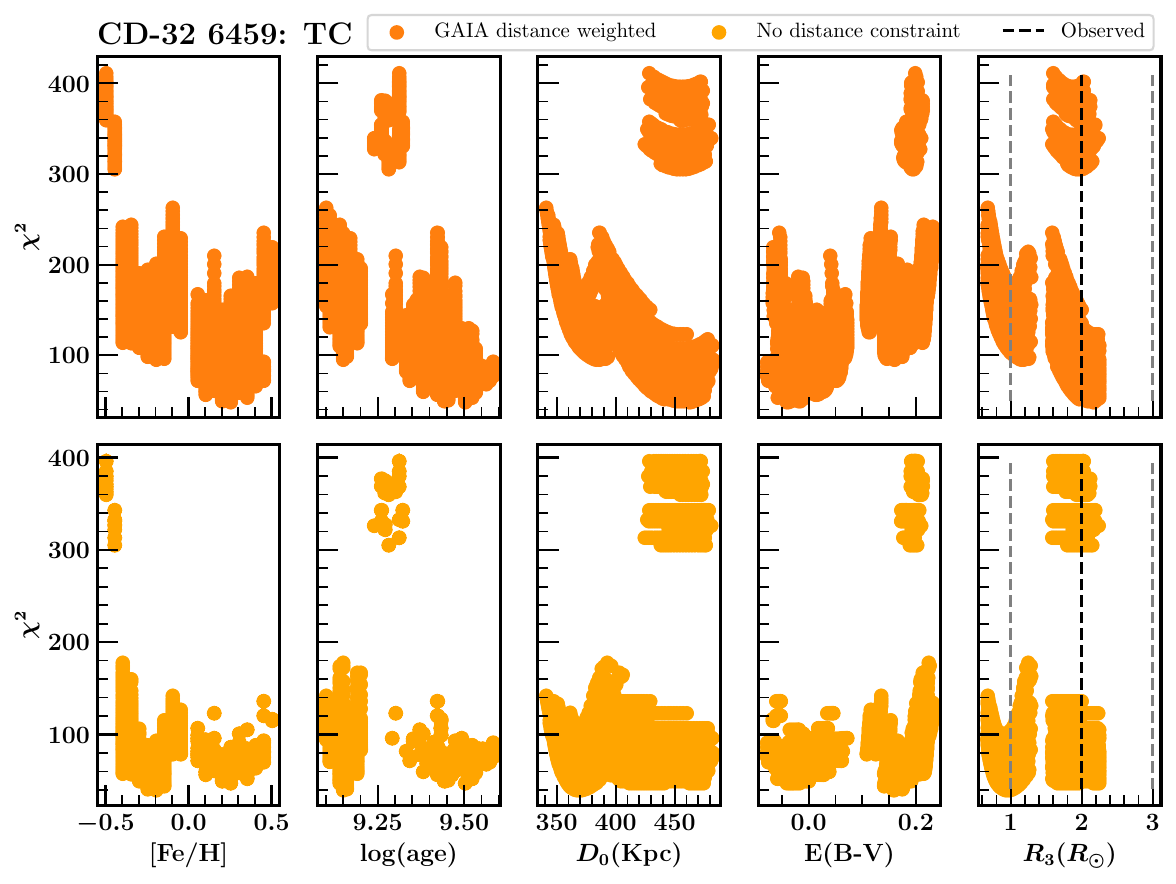}
    \includegraphics[width=0.46\textwidth]{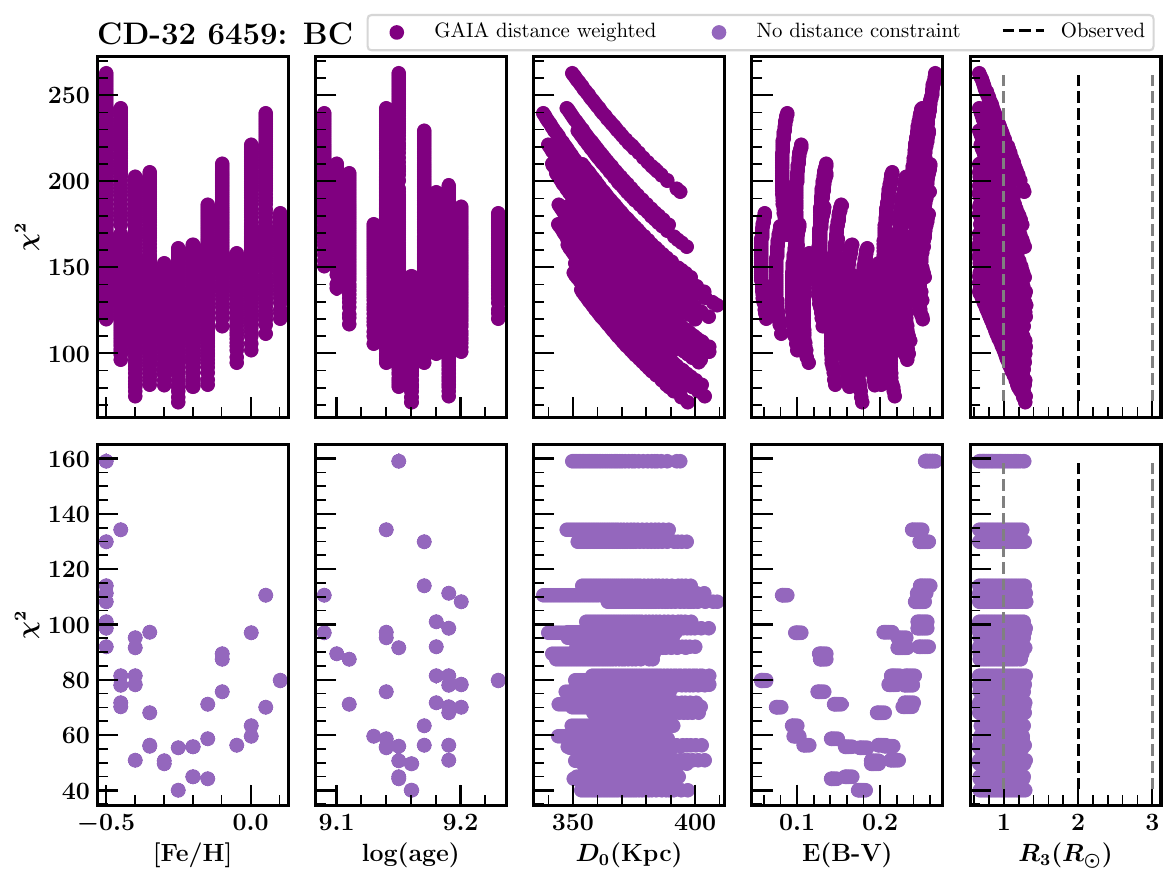} \\  
    \centering
    \includegraphics[width=0.6\textwidth]{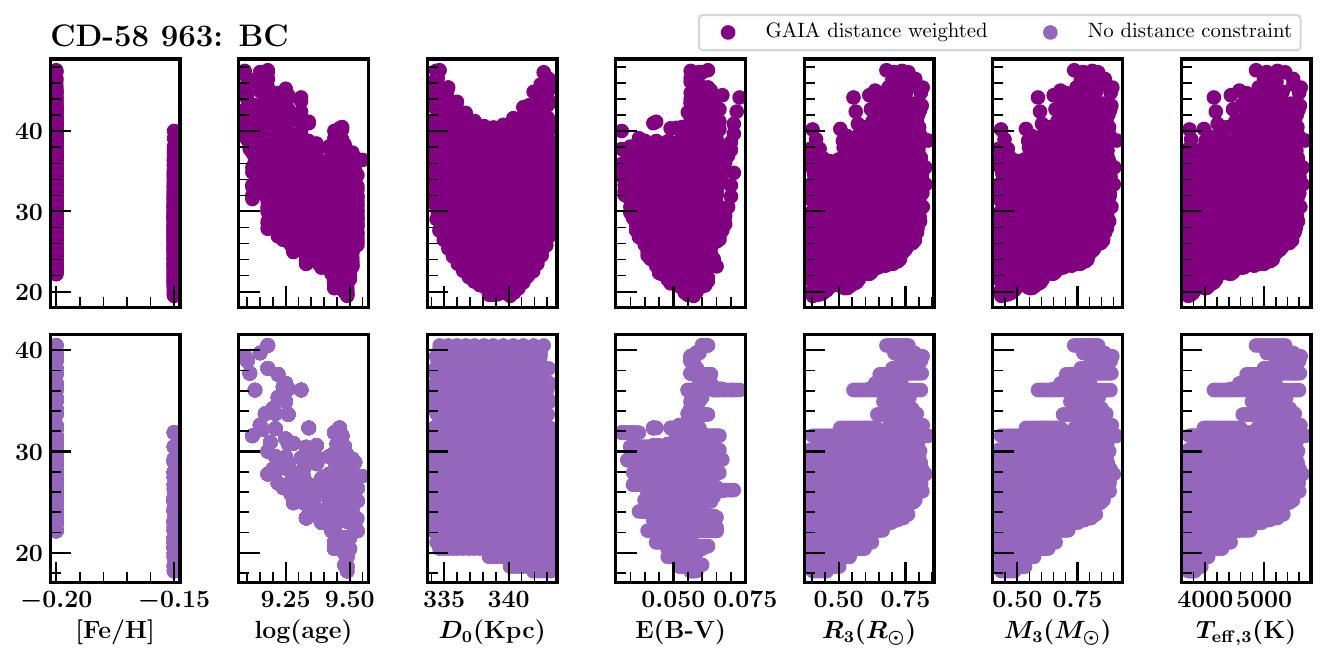} \\
        \includegraphics[width=0.6\textwidth]{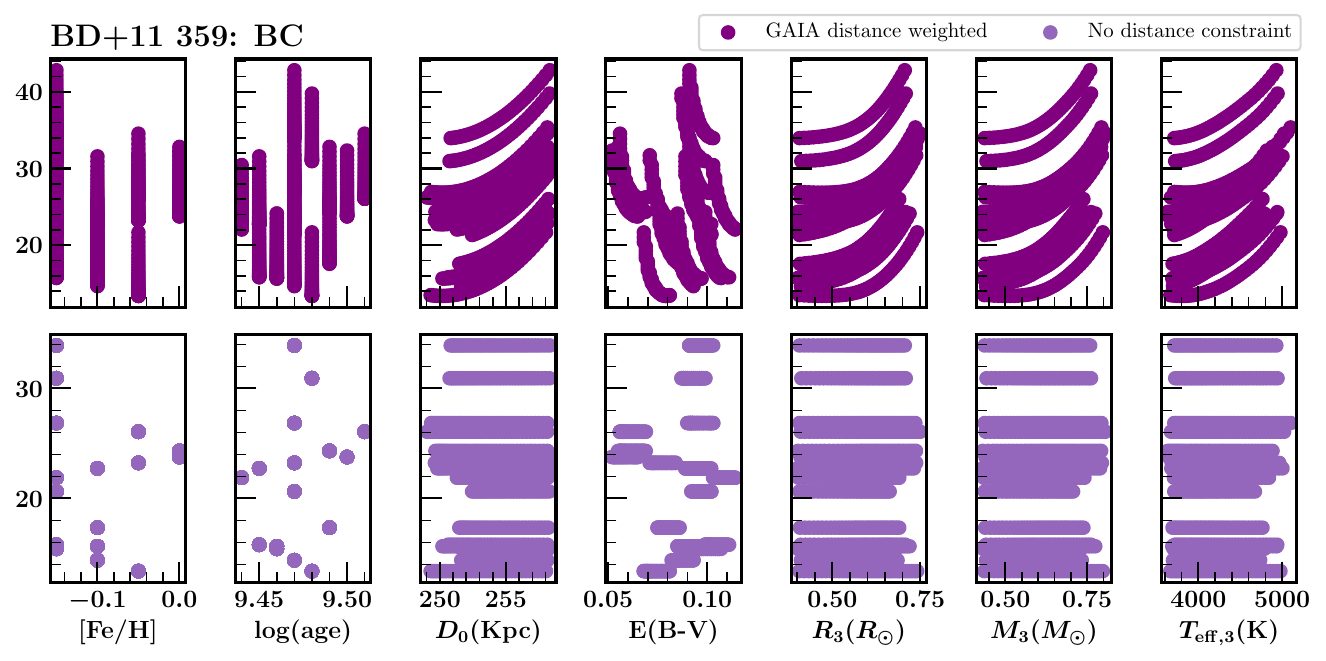}

    \caption{\textsc{isofitter} $\chi^2$ grids for specific parameters for different constraints. Orange denotes grids calculated with constraints from all stars while purple denotes grids calculated using only the eclipsing pair.}
    \label{fig:isofitterruns}
\end{figure*}

The isochrone fitting code \textsc{isofitter} calculates grids of $\chi^2$ using models from MIST and the measured parameters. In \autoref{fig:isofitterruns} we show the $\chi^2$ map for isochrone metallicity ([Fe/H]), log of the system age (log(age)), reddening-free distance ($D_0$), extinction (as $E(B-V)$) and estimated radius of the tertiary ($R_3$).  The top panel for each star shows grids where the $\chi^2$ was weighted with {\it Gaia} distance and the lower panels were free of this weight. For ST3 systems, we calculate separate grids using three-star constraints (TC; \autoref{fig:isofitterruns}-left) and binary constraints (BC; \autoref{fig:isofitterruns}-right). For the case of ST2 systems, we use only BC and we show the expected mass of the tertiary ($M_3$) as well.

\section{Table of detailed CHT solutions}
There are 48 systems in the literature with a complete set of orbital, stellar and atmospheric parameters of all stars in a CHT. We list all these systems in \autoref{tab:CHTlit}. The systems are characterised according to their method of estimation of orbital parameters, metallicity, and ages.

\begin{table*}[]
    \centering
  \caption{CHTs with detailed parameters, and their sources.}
    \label{tab:CHTlit}
\begin{tabular}{lrrr}
\hline
  \multicolumn{1}{c}{System} &
  \multicolumn{1}{c}{$P_1$} &
  \multicolumn{1}{c}{$P_2$} &
  \multicolumn{1}{c}{Source} \\
\hline
\hline
\multicolumn{4}{c}{Triply eclipsing systems} \\
\hline
  HD 181068 & 0.9057 & 45.471 & \cite{Derekas2011,Borko2013}\\ 
   HD144548$^{a,b}$ & 1.6278 & 33.945 & \cite{Alonso,hd144noncoeval}\\ 
     EPIC 249432662 & 8.1941 & 188.379 & \cite{Borko2019}\\ 
    TIC 209409435 & 5.7175 & 121.872 & \cite{Borko2020}\\
      TIC 278825952 & 4.7810 & 235.550 & \cite{Mitnyan2020}\\
    KIC 7955301 & 15.3183 & 209.760 & \cite{Gaulme2022}\\
     TIC 388459317 & 2.1849 & 89.0312 & \cite{BorkoF2022}\\
  TIC 193993801 & 1.4308 & 49.4020 & \cite{BorkoF2022}\\
    TIC 52041148 & 1.7872 & 177.8620 & \cite{BorkoF2022}\\ 
      TIC 242132789 & 5.1287 & 42.0317 & \cite{Rapp2022}\\ 
  TIC 42565581$^{a,b}$ & 1.8235 & 123.5467 & \cite{Rapp2022}\\ 
TIC 54060695 & 1.0608 & 60.7759 & \cite{Rapp2022}\\ 
  TIC 37743815 & 0.9069 & 68.7998 & \cite{Rapp2022}\\
    TIC 456194776 & 1.7193 & 93.9150 & \cite{Rapp2022}\\
      TIC 178010808 & 0.8264 & 69.0830 & \cite{Rapp2022}\\
     KOI-126  & 1.7222 & 34.0170 & \cite{yenawinne2022} \\
  TIC 294803663 & 2.2456 & 153.4260 & \cite{RappMy2023}\\
      TIC 99013269 & 6.5344 & 604.2425 & \cite{RappMy2023}\\
  TIC 332521671 & 1.2479 & 48.5848 & \cite{RappMy2023}\\
  TIC 47151245 & 1.2025 & 284.3740 & \cite{RappMy2023}\\
        TIC 276162169 & 2.5498 & 117.2700 & \cite{RappMy2023}\\
  TIC 280883908 & 5.2418 & 184.5980 & \cite{RappMy2023}\\
  TIC 229785001 & 0.9298 & 165.3700 & \cite{RappMy2023}\\
  TIC 356324779 & 3.4717 & 87.0920 & \cite{RappMy2023}\\
  TIC 81525800 & 1.6131 & 49.7500 & \cite{RappMy2023}\\
    TIC 298714297 & 1.0729 & 117.2400 & \cite{Czava2023}\\
  TIC 66893949 & 16.2900 & 386.4000 & \cite{Czava2023}\\
    TIC 88206187 & 1.1846 & 52.9220 & \cite{Czava2023}\\
    KIC 7668648 & 27.7963 & 208.0050 & \cite{orosz2023} \\
    TIC 650024463 & 7.1978 & 108.7251 & \cite{Rapp2024}\\
    TIC 133771812 & 12.3339 & 243.8900 & \cite{Rapp2024}\\
      TIC 185615681 & 2.3180 & 56.0666 & \cite{Rapp2024}\\
        TIC 176713425 & 1.8951 & 52.9430 & \cite{Rapp2024}\\
          TIC 287756035 & 2.0822 & 367.9230 & \cite{Rapp2024}\\
     TIC 321978218$^{a,b}$ & 0.5703 & 57.5354 & \cite{Rapp2024}\\
       TIC 323486857 & 0.8850 & 41.4268 & \cite{Rapp2024}\\  
\hline
\multicolumn{4}{c}{Eclipse depth variation  + eclipse timing variation} \\
\hline
  KIC 5731312 & 7.9465 & 917.0000 & \cite{BorkoS2022}\\
  KIC 5653126 & 38.4482 & 971.3900 & \cite{BorkoS2022}\\
  KIC 8023317 & 16.5754 & 605.4000 & \cite{BorkoS2022}\\
   KIC 6964043$^c$ & 10.6979 & 239.2519 & \cite{BorkoS2022}\\
  KIC 10319590 &  21.26551 &  455.7000 &  \cite{orosz2023}\\
  KIC 5771589 & 10.6791 & 113.8720 & \cite{BorMit24}\\
\hline
\multicolumn{4}{c}{Eclipse timing variation} \\
\hline
TIC 219885468 & 7.5128 & 111.5498 & \cite{BorMit24}\\
KIC 9714358 & 6.4708 & 104.0830 & \cite{BorMit24}\\
  GSC 08814-0102$^b$ & 0.7024 & 245.0000 & \cite{helminiak_gsc,moharanasol}\\
\hline
\multicolumn{4}{c}{Spectroscopy  of doubly eclipsing systems} \\
\hline
  DY Lyncis  & 1.3132 & 281.1800 & \cite{dimitrovdylyn}\\
  BD+44 2258 & 3.4726 & 254.8400 & \cite{moharanacht}\\
  KIC 6525196 & 3.4206 & 418.0000 & \cite{krishides2017,moharanacht}\\
\hline
\hline
\end{tabular} \\
$^a$ Tertiary and binary have different age estimates. $^b$ log(age) and [Fe/H] from the inner binary. $^c$ Also shows triple eclipses.
\\
\end{table*}
\end{appendix}
\end{document}